\documentclass[twocolumn,aps,superscriptaddress,showpacs,nofootinbib,floatfix]{revtex4-1}
\usepackage{epsfig,bm}
\usepackage{graphics}
\usepackage{amsmath}
\usepackage[normalem]{ulem}  
\usepackage[dvips]{color} 

\renewcommand\sout{\bgroup \color{red} \ULdepth=-.5ex \ULset}

\begin{document}


\title{Chiral symmetry restoration in heavy-ion collisions at intermediate energies}

\author{A. Palmese}
\affiliation{Institut f\"{u}r Theoretische Physik, Universit\"{a}t Giessen, Germany}

\author{W. Cassing}
\affiliation{Institut f\"{u}r Theoretische Physik, Universit\"{a}t Giessen, Germany}

\author{E. Seifert}
\affiliation{Institut f\"{u}r Theoretische Physik, Universit\"{a}t Giessen, Germany}

\author{T. Steinert}
\affiliation{Institut f\"{u}r Theoretische Physik, Universit\"{a}t Giessen, Germany}

\author{P. Moreau}
\affiliation{Institute for Theoretical Physics, Johann Wolfgang Goethe Universit\"{a}t, Frankfurt am Main, Germany}

\author{E. L. Bratkovskaya}
\affiliation{Institute for Theoretical Physics, Johann Wolfgang Goethe Universit\"{a}t, Frankfurt am Main, Germany}
\affiliation{GSI Helmholtzzentrum f\"{u}r Schwerionenforschung GmbH, Darmstadt, Germany}
\begin{abstract}
We study the effect of the chiral symmetry restoration (CSR) on
heavy-ion collisions observables in the energy range
$\sqrt{s_{NN}}$=3--20\,GeV within the Parton-Hadron-String Dynamics
(PHSD) transport approach. The PHSD includes the deconfinement phase
transition as well as essential aspects of CSR in the dense and hot
hadronic medium, which are incorporated in the Schwinger mechanism
for the hadronic particle production. We adopt different
parametrizations of the nuclear equation of state from the
non-linear $\sigma-\omega$ model, which enter in the computation of
the quark scalar density for the CSR mechanism, in order to estimate
the uncertainty in our calculations. For the pion-nucleon
$\Sigma$-term we adopt $\Sigma_\pi \approx$ 45 MeV which corresponds
to some 'world average'. Our systematic studies show that chiral
symmetry restoration plays a crucial role in the description of
heavy-ion collisions at $\sqrt{s_{NN}}$=3--20\,GeV, realizing an
increase of the hadronic particle production in the strangeness
sector with respect to the non-strange one. We identify particle
abundances and rapidity spectra to be suitable probes in order to
extract information about CSR, while transverse mass spectra are
less sensitive. Our results provide a microscopic explanation for
the "horn" structure in the excitation function of the $K^+/\pi^+$
ratio: the CSR in the hadronic phase produces the steep increase of
this particle ratio up to $\sqrt{s_{NN}} \approx$ 7 GeV, while the
drop at higher energies is associated to the appearance of a
deconfined partonic medium. Furthermore, the
appearance/disappearance of the 'horn'-structure is investigated as
a function of the system size and collision centrality. We close
this work by an analysis of strangeness production in the ($T,
\mu_B$)-plane (as extracted from the PHSD for central Au+Au
collisions) and discuss the perspectives to identify a possible
critical point in the phase diagram.

\end{abstract}


\maketitle

\section{Introduction}
\renewcommand{\thefootnote}{\fnsymbol{footnote}}
The main goal of heavy-ion collision (HIC) physics is the study of
the behavior of nuclear matter at high temperatures and/or high
densities. In particular, the major interest is the investigation of
the nuclear matter phase diagram as a function of temperature and
baryon chemical potential. According to Quantum-Chromo-Dynamics
(QCD), at large temperatures and densities the hadrons cannot
survive anymore as bound states and they dissolve forming the
so-called Quark-Gluon-Plasma (QGP). At Relativistic Heavy-Ion
Collider (RHIC) energies for the first time the creation of a QGP,
identified as an almost perfect fluid \cite{Romatschke}, has been
proven experimentally. Actually, the properties of this deconfined
state of  matter are still debated as well as the phase boundaries
to the hadronic phase. In order to shed some light on this issue,
many heavy-ion experiments are performed at the Super-Proton
Synchrotron (SPS),  RHIC, the Large Hadron Collider (LHC) and will
be performed at the future Facility for Antiproton and Ion Research
(FAIR) as well as the Nuclotron-based Ion Collider fAcility (NICA).
The crucial challenge is to identify in the final particle
distributions those signatures, which allow to disentangle the QGP
contribution that is impossible to observe directly or
independently.

The strange particle production has always been suggested as one of
the most sensitive observables that could spot out the creation of a
QGP during the early stages  of a HIC. The earliest suggested
signature is the strangeness enhancement in A+A collisions with
respect to elementary p+p collisions \cite{Rafelski,stock}. Later
on, Gazdzicki and  Gorenstein \cite{SMES} proposed that a sharp rise
and drop in the excitation function of the $K^+/\pi^+$ ratio (so
called "horn") should show up due to the appearance of a QGP phase
at a center-of-mass energy $\sqrt{s_{NN}}\sim 7\,$GeV
\footnote[2]{In this work we adopt natural units, hence $\hbar=c=1$.}.
Several statistical models \cite{Cleymans:2004hj,ABS06,Bugaev} have succeeded in reproducing
the trend of the experimental observation of the $K^+/\pi^+$ ratio
and other strange to non-strange particle ratios, but they can
provide only a statistical description of the heavy-ion collision
process. On the other hand there was no conclusive interpretation of
the "horn" from dynamical approaches for HIC, like microscopic
transport models \cite{Jgeiss,Brat04,BratPRL}. Only recently, the
Parton-Hadron-String Dynamics (PHSD), a transport approach
describing HIC on the basis of partonic, hadronic and string
degrees-of-freedom, obtained a striking improvement on this issue
when including chiral symmetry restoration (CSR) in the string decay
for hadronic particle production \cite{PHSD_CSR}. Within the PHSD
approach it has been suggested that the horn feature emerges in the
energy-dependence of the $K^+/\pi^+$ ratio, both due to CSR, which
is responsible for the rise at low energies, and to the appearance
of a deconfined partonic medium at higher energies, which is
responsible for the drop at top SPS energies.

Apart from deconfinement the chiral symmetry restoration addresses
another aspect of the QCD phase diagram in the ($T, \mu_B$)-plane as
an additional  transition between a phase with broken and a phase
with restored chiral symmetry. As in case of the QCD deconfinement
phase transition, the boundaries of the CSR phase transition line
are not well known. Lattice QCD (lQCD) calculations show that at
vanishing baryon chemical potential $\mu_B$=0 the CSR takes place at
roughly the same critical temperature and energy density as the
deconfinement phase transition which is a crossover. At finite
baryon chemical potential lQCD calculations cannot be performed due
to the sign problem and one must rely on effective models (or
extrapolations) in order to study the QCD phase transitions
\cite{CBMbook,Ch1,Ch2,Ch3,BJS}. Different models support the idea
that at finite chemical potential a partially restored phase is
achieved before the deconfinement occurs
\cite{Rob1,Nambu1961,Klevansky1992}. In order to distinguish the two
phases of such a transition, effective models use the scalar quark
condensate $\langle \bar q q \rangle$ as an order parameter. As the
baryon density and temperature increase, the scalar quark condensate
$\langle \bar q q \rangle$ is expected to decrease from a
non-vanishing value in the vacuum to $\langle \bar q q
\rangle\approx\,0$ which corresponds to CSR. Since $\langle \bar q q
\rangle$ is not a measurable quantity, it is crucial to determine
experimental observables which are sensitive to this feature. Since
long the dilepton spectroscopy has been in the focus in this respect
since in a chirally restored phase the spectral functions of the the
$\rho$- and the $a_1$-meson should become identical. However, no
clear evidence has been achieved so far \cite{dileptons}. On the
other hand, the strangeness production at Alternating-Gradient
Synchrotron (AGS) and lower SPS energies was suggested to be a
signature of CSR \cite{PHSD_CSR}.

In this work we will extend the analysis in Ref.
\cite{PHSD_CSR} and perform a systematic study within the PHSD
approach on effects of the CSR on final particle distributions in
HICs in the energy range $\sqrt{s_{NN}}=3-20\,$GeV. The PHSD  is
particularly suited to this aim since it includes both hadronic and
partonic degrees-of-freedom \cite{PHSD} and  has been extensively
used to describe HIC observables from SPS to LHC energies
\cite{PHSDrhic,Linnyk,Volo}.

This study is organized as follows: In Sec. II we briefly recall the
basic ingredients of the PHSD approach, while in Sec. III we
illustrate the string fragmentation included in PHSD and the
most recent extension of this particle production mechanism to
incorporate essential aspects of chiral symmetry restoration.
Whereas in Ref. \cite{PHSD_CSR} the focus has been on the effects of
CSR and in particular of the partonic phase, we here also discuss
the sensitivity of our results on different parameter settings for
the nuclear Equation of State (EoS) within the non-linear
$\sigma-\omega$ model and examine the role of three-meson channels
for strangeness production ($3M\leftrightarrow B\bar B$). In Sec. IV
we present the calculated results from PHSD -- with and without the
inclusion of CSR -- for the particle rapidity spectra incorporating 
different nuclear equations of state in order to
investigate the uncertainties of our approach. In addition to Ref.
\cite{PHSD_CSR} we evaluate the transverse mass spectra for protons,
pions, kaons and antikaons from central collisions of heavy systems
in the energy range from 2 to 158 A GeV  with a special focus on the
strange particle production. The excitation functions for the ratios
$K^+/\pi^+$, $K^-/\pi^-$ and $(\Lambda + \Sigma^0)/\pi$ - together
with their uncertainties - complete Sec. IV where a comparison with
experimental data is performed wherever possible. In Sec. V we
explore further new aspects of CSR in heavy-ion collisions, i.e. the
system size and the centrality dependence of strange particle yields
and ratios. In Sec. VI we focus on the equilibrium stages of the
time evolution of central Au+Au collisions in connection to
strangeness production in the ($T,\mu_B$)-plane and discuss the
perspectives to identify a critical point in the phase diagram. We
conclude this work with a summary in Sec.VII.

\section{Reminder of the PHSD transport approach}\label{PHSD}
The Parton-Hadron-String Dynamics (PHSD) is a microscopic covariant
dynamical approach for strongly interacting systems in and out-of
equilibrium \cite{PHSD,PHSDrhic}. It is a transport approach which
goes beyond the quasi-particle approximation, since it is based on
Kadanoff-Baym equations for the Green`s functions in phase-space
representation in first-order gradient expansion
\cite{Kadanoff,Cassing:2008nn}. Including both a hadronic and a
partonic phase as well as a transition between the effective
degrees-of-freedom, PHSD is capable to describe the full time
evolution of a relativistic heavy-ion collision. The theoretical
description of the partonic degrees-of-freedom (quarks and gluons)
is realized in line with the Dynamical-Quasi-Particle Model (DQPM)
\cite{Cassing:2008nn} which  reproduces lQCD results in
thermodynamical equilibrium and provides the properties of the
partons, i.e. masses and widths in their spectral functions. In
equilibrium the PHSD reproduces the partonic transport coefficients
such as shear and bulk viscosities or the electric conductivity from
lQCD calculations as well \cite{PHSDrev}.

An actual nucleus-nucleus collision in PHSD follows the following steps:
\begin{itemize}
\item Primary hard scatterings between nucleons take place and produce excited color-singlet states, denoted by "strings" (as described within the FRITIOF Lund model \cite{FRITIOF} and realized also in PYTHIA 6.4 \cite{Sjostrand:2006za}). These strings decay into "prehadrons" with a formation time $\tau_f\sim 0.8\,$fm/c and "leading hadrons", which originate from the string ends and can re-interact with hadrons almost instantly with reduced cross-sections (according to the constituent quark number).
\item In case the local energy density is above the critical value of $\epsilon_c\sim0.5\,$GeV/fm$^3$, the deconfinement is implemented by dissolving the newly produced hadrons into the massive colored quarks/antiquarks and mean-field energy from the DQPM.
\item  Within the QGP phase, the partons (quarks, antiquarks and gluons) scatter and propagate in a self-generated mean-field potential. They are described as off-shell quasi-particles with temperature-dependent masses and widths, which are given by the DQPM.
\item The expansion of the system is associated to a decrease of the local energy density and, once the local energy density becomes close to or lower than $\epsilon_c$, the massive colored off-shell quarks and antiquarks hadronize to colorless off-shell mesons and baryons. The hadronization process is defined by covariant transition rates and fulfills the energy-momentum and quantum number conservation in each event.
\item In the hadronic corona as well as in the late hadronic phase, the particles are still interacting and propagating. Elastic and inelastic collisions between baryons,
mesons and resonances are implemented in PHSD and the corresponding backward reactions are included through detailed balance for all channels.
\end{itemize}

We further note that the pure hadronic phase in PHSD is equivalent
to the Hadron-Strings Dynamics (HSD) model \cite{HSD}. Accordingly,
the comparison between PHSD and HSD calculations allows us to
disentangle the role of the QGP phase in heavy-ion collisions. Even
at low center-of-mass energies ($\sqrt{s_{NN}}\sim3\,$GeV) the PHSD
and the HSD results slightly differ due to the appearance of  QGP
'droplets' in central cells of the system, which are characterized
by high baryon and energy densities.

The PHSD approach has been tested for different colliding systems
(p+p, p+A, A+A) in a wide range of bombarding energy, from AGS to
LHC energies, and has been able to describe a large number of
experimental observables, such as charged particle spectra,
collective flow coefficients $v_n$ as well as electromagnetic probes
such as photons and dileptons \cite{PHSDrev}. More recently, in Ref.
\cite{PHSD_CSR}, it has been also shown to provide a microscopic
description of the maximum in the $K^+/\pi^+$ ratio in central
nucleus-nucleus collisions. In the latter work we also found that
the inclusion of CSR in the hadronic sector via string decay is
crucial in order to reproduce the strangeness enhancement at AGS and
lower SPS energies.

\section{String fragmentation in PHSD}

\subsection{Basic concepts}
The string formation and decay represents the dominant particle production mechanism in nucleus-nucleus collisions for bombarding energies from $2\,$AGeV to $160\,$AGeV.
In PHSD, the primary hard scatterings between nucleons are described by string formation and decay in the FRITIOF Lund model \cite{FRITIOF}.
A string is an excited color-singlet state, which is composed of two string ends corresponding to the leading constituent quarks of the colliding hadrons and a color flux tube in between. As the string ends recede, virtual $q\bar q$ or $qq\bar q\bar q$ pairs are produced in the uniform color field, causing the breaking of the string. Finally, the string decays into mesons or baryon-antibaryon pairs with formation time $\tau_f\sim 0.8\,$fm/c (in the rest-frame of the string).
\\In the string decay, the flavor of the produced quarks is determined via the Schwinger formula \cite{Schwinger,Sjostrand:2006za}, which defines the production probability of massive $s\bar s$ pairs with respect to light flavor production $(u\bar u,d\bar d)$ pairs:
\begin{equation}
\label{Schwinger-formula}
\frac{P(s\bar s)}{P(u\bar u)}=\frac{P(s\bar s)}{P(d\bar d)}=\gamma_s=\exp\Bigl(-\pi\frac{m_s^2-m_{u,d}^2}{2\kappa}\Bigr)\,,
\end{equation}
with $\kappa\approx 0.176\,$GeV$^2$ representing the string tension and $m_{u,d,s}$ denoting the constituent quark masses for strange and light quarks.
For the constituent quark masses $m_u\approx0.35\,$GeV and $m_s\approx0.5\,$GeV in the vacuum, the production of strange quarks is suppressed by a factor of $\gamma_s\approx 0.3$ with respect to the light quarks, which is the default setting in the FRITIOF routines.
The relative production factors in PHSD/HSD have been  readjusted in 1998 as follows \cite {Jgeiss}:
\begin{eqnarray}
\label{schwinger_factors}
u:d:s:uu = \left\{
\begin{array}{ll}
1:1:0.3:0.07 & \mbox{at SPS to RHIC; } \\ 1:1:0.4:0.07 &
\mbox{at AGS energies.}
\end{array}
\right.
\end{eqnarray}
The probability ratio $\gamma_s$ has been increased to $0.4$ at AGS
energies in order to correctly reproduce the strangeness yield for
p+Be collisions at AGS energies \cite {Jgeiss}. A smooth transition
between the two values of $\gamma_s$ is ensured by a linear
interpolation as a function of the center-of-mass energy $\sqrt{s}$.
\\A further ingredient to fix the rapidity distribution for the newly produced hadrons is the fraction of energy and momentum that they acquire from the decaying string. This is defined by the fragmentation function $f(x,m_T)$, which is the probability distribution for a hadron with transverse mass $m_T$ to be produced with an energy-momentum fraction $x$ from the fragmenting string:
\begin{equation} \label{frag0}
f(x,m_T)\approx \frac{1}{x}(1-x^a)\exp(-b\,m_T^2/x)\,,
\end{equation}
where $a=0.23$ and $b=0.34\,$GeV$^{-2}$ as reliable settings for p+p
and p+A collisions. As becomes evident from Eq. (\ref{frag0}) the
meson $m_T$-scaling from string decay is included by default.

\subsection{Modeling of the chiral symmetry restoration}
In Ref. \cite{PHSD_CSR} the PHSD has been extended to include CSR in
the string decay in a hadronic environment of finite baryon and
meson density. Here we recall the main aspect of this extension
which is based on the Hellman-Feynman theorem for the scalar quark
condensate \cite{Cohen92}. Accordingly, a linear decrease of the
scalar quark condensate $\langle{\bar q} q\rangle$ -- which is
nonvanishing in the vacuum due to a spontaneous breaking of chiral
symmetry -- is expected with baryon density $\rho_B$ towards a
chiral symmetric phase characterized by $\langle\bar{q} q\rangle
\approx$  0 \cite{Weise,Birse}. This decrease of the scalar quark
condensate is expected also to lead to a change of the hadron
properties with density and temperature, i.e. in a chirally restored
phase the vector and axial vector currents should become equal
\cite{Kochr,Zahed,GEB,GEB1,Koch}; the latter implies that e.g. the
$\rho$ and $a_1$ spectral functions should become identical (as
addressed above in the context of dilepton production). Since the
scalar quark  condensate $\langle\bar{q} q\rangle$ is not a direct
observable, its manifestations should also be found indirectly in
different hadronic abundances and spectra or particle ratios like
$K^+/\pi^+$, $(\Lambda+\Sigma^0)/\pi^-$ etc. as advocated in Ref.
\cite{PHSD_CSR}.

In leading order the scalar quark condensate $\langle \bar q q
\rangle$ can be evaluated in a dynamical calculation as follows
\cite{Toneev98}:
\begin{equation}
\frac{\langle\bar{q} q\rangle}{\langle\bar{q} q\rangle_V} = 1 - \frac{\Sigma_\pi}{f_\pi^2 m_\pi^2}\rho_S - \sum\limits_h{\sigma_h \rho_S^h \over f_\pi^2
m_\pi^2}\,,
\label{scalar_condensate}
\end{equation}
where $\sigma_h$ stands for the $\sigma$-commutator of the relevant
mesons $h$, $\langle{\bar q} q\rangle_V$ represents the vacuum
condensate, $\Sigma_\pi \approx45\,$MeV is the pion-nucleon
$\Sigma$-term and $f_\pi$ and $m_\pi$ are the pion decay constant
and pion mass, respectively. Note, however, that the value of
$\Sigma_{\pi} $ is not so accurately known; a recent analysis points
towards a larger value of $\Sigma_{\pi} \approx 59$ MeV
\cite{Alarcon,Meissner} while actual lQCD results \cite{Sternberg}
suggest a substantially lower value. Accordingly, our following
calculations - based on $\Sigma_{\pi} = 45$ MeV - have to be taken
with some care although in some sense it represents a 'world
average' (cf. Fig. 3 in Ref. \cite{sigmapi}).
%
According to the light quark content, the $\Sigma$-term for hyperons
is decreased by a factor of 2/3 for $\Lambda$ and $\Sigma$ hyperons
and by a factor of 1/3 for $\Xi$ baryons. Furthermore, for mesons
made out of light quarks and antiquarks, we use $\sigma_h =
m_\pi/2$,  whereas for mesons with a strange (antistrange) quark we
consider $\sigma_h = m_\pi/4$. We mention here that improved results
for the sigma-commutator for kaons can be obtained from chiral
perturbation theory as in Ref. \cite{Bordes} and alternative
assumptions for non-pseudoscalar mesons can be employed as e.g.
suggested by Cohen et al. in Ref. \cite{Cohen92}. In view of the
subleading contributions of these mesons to the ratio in Eq.
(\ref{scalar_condensate}) we keep the simple estimates noted above
for our present study and look forward to a clarification of the
present tension between the results from lQCD and dispersive
approaches \cite{Meissner2}.

In Eq. (\ref{scalar_condensate}), the quantities $\rho_S$ and
$\rho_S^h$ denote the nucleon scalar density and the scalar density
for a meson of type $h$, respectively. The scalar density of mesons
$h$ is evaluated in the independent-particle approximation as:
\begin{equation}
\label{rhos_mesons} \rho_S^h(x) = \frac{(2s+1) (2\tau+1)}{(2\pi)^3} \int \mathrm{d}^3 p \frac{m_h}{\sqrt{{\bf p}^2 + m_h^2}} f_h(x,{\bf
p})\,,
\end{equation}
where $f_h(x,{\bf p})$ denotes the meson phase-space distribution
($x=({\bf r}, t)$) and $s,\tau$ refer to the discrete spin and
isospin quantum numbers, respectively. Moreover, the vacuum scalar
condensate  $ \langle\bar{q} q\rangle_V = \langle\bar{u} u\rangle_V
+\langle\bar{d} d\rangle_V \approx 2 \langle\bar{u} u\rangle_V$ can
be computed according to the Gell-Mann-Oakes-Renner (GOR) relation
\cite{GOR,Cohen},
\begin{equation} f_\pi^2 m_\pi^2
= - \frac{1}{2} (m_u^0 + m_d^0) \langle\bar{q}
q\rangle_V\,,\end{equation} and gives $\langle\bar{q} q\rangle_V
\approx- 3.2\,$fm$^{-3}$ for the bare quark masses $m_u^0 = m_d^0
\approx7\,$MeV. Finally, in Eq. (\ref{scalar_condensate}) the
nucleon scalar density $\rho_S$ has to be determined in a suitable
model with interacting degrees-of-freedom in order to match our
knowledge on the nuclear EoS at low temperature and finite density.
A proper (and widely used) approach is the non-linear
$\sigma-\omega$ model for nuclear matter where $\rho_S$ is defined
as:
\begin{equation}
\label{rhos_nucleon} \rho_S(x) = \frac{g_n}{(2\pi)^3} \int \mathrm{d}^3
 p \frac{m^*_N}{\sqrt{{\bf p}_N^{*2} + m_N^{*2}}} f_N(x,{\bf
p})\,,
\end{equation}
where $m^*_N$ and $p^*_N$ denote the effective mass and momentum,
respectively, and $f_N(x,{\bf p})$ the phase-space occupation of a
nucleon while the degeneracy factor is $g_n$=4. In fact, in the
non-linear $\sigma-\omega$ model the nucleon mass is modified due to
the scalar interaction with the medium:
\begin{equation}
\label{mass} m_N^*(x)= m_N^V- g_s \sigma(x)\,,
\end{equation}
where $m_N^V$ denotes the nucleon mass in vacuum and $\sigma(x)$ is
the scalar field which mediates the interaction between the nucleons
and the medium with the coupling $g_s$. In order to calculate
$\rho_S$, we need to determine the value of the scalar field
$\sigma(x)$ at each space-time point $x$. This is done via the
non-linear gap equation \cite{Boguta,Lang}:
\begin{equation} \label{gap_equation}
m_\sigma^2 \sigma(x) + B \sigma^2(x) + C \sigma^3(x) = g_s \rho_S(x)
\end{equation} $$=
 g_s d \int \frac{\mathrm{d}^3 p}{(2 \pi)^3}
\frac{m_N^*(x)}{\sqrt{{\bf p}^2+m_N^{*2}}} f_N(x,{\bf p})\,,
$$ since for matter at rest we have ${\bf p}^*= {\bf p}$.
In Eq. (\ref{gap_equation}) the self-interaction of the
$\sigma$-field is included up to the forth order. The parameters
$g_s, m_\sigma, B, C$ are fixed in order to reproduce the values of
the nuclear matter quantities at saturation, i.e. the saturation
density, the binding energy per nucleon, the compression modulus,
and the effective nucleon mass. Actually, there are different sets
for these quantities that lead to slightly different saturation
properties. We defer a discussion on the uncertainties of our
results to the following section \ref{EOS}.

The main idea in Ref. \cite{PHSD_CSR} is to consider effective
masses for the dressed quarks in the Schwinger formula
(\ref{Schwinger-formula}) for the string decay in a hot and dense
medium. The effective quark masses can be expressed in terms of a
scalar coupling to the quark condensate $\langle \bar q q \rangle$
in first order as follows:
\begin{equation}   \label{effective_masses}
m_s^* = m_s^0 + (m_s^V-m_s^0) \frac{\langle \bar q q \rangle}{\langle \bar q q \rangle_V}\,, \end{equation}
\begin{equation}   \label{mus}
m_q^* = m_q^0 + (m_q^V-m_q^0)  \frac{\langle \bar q q \rangle}{\langle \bar q q \rangle_V}\,,
\end{equation}
with $m_s^0 \approx 100\,$MeV and $m_q^0 \approx 7\,$MeV for the
bare quark masses. In Eq. (\ref{effective_masses})  the effective
masses decrease from the vacuum values with decreasing scalar
condensate $\langle \bar q q \rangle$ to the constituent masses.
This adaptation of the Schwinger formula in case of a hot and dense
medium implies a modification of the flavor production factors in
Eq. (\ref{schwinger_factors}). In an actual nucleus-nucleus
collision, PHSD incorporates a dynamical calculation of all these
features for each cell in space-time:
\begin{itemize}
\item the scalar density $\rho_S$ is determined by solving the gap equation (\ref{gap_equation}) for the $\sigma$-field;
\item the scalar condensate $\langle \bar q q \rangle$ is then computed via Eq. (\ref{scalar_condensate});
\item the effective masses $m_q^*,m_s^*$ are calculated according to Eqs. (\ref{effective_masses}), (\ref{mus}) and plugged in the Schwinger formula (\ref{Schwinger-formula}) in order to compute the flavor production ratios for the string decay.
\end{itemize}
We stress that, once the nucleon scalar density $\rho_S$ and
$\Sigma_{\pi}$ are fixed, there is no additional 'parameter' in the
PHSD3.3 compared to the previous version PHSD3.2 that has been
employed for a couple years for the analysis of relativistic
heavy-ion reactions \cite{PHSDrev}.

\subsection{Dependence on the nuclear equation of state}
\label{EOS}
\begin{figure}[t!]
\centering
\includegraphics[width=0.46\textwidth]{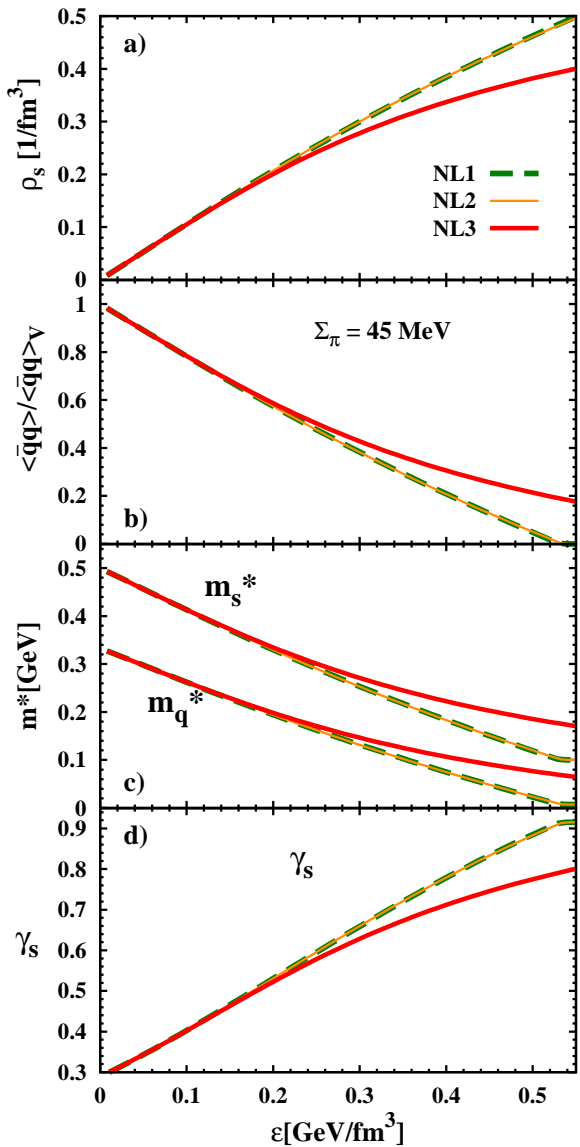}
\caption{(Color online) The nucleon scalar density $\rho_S$ (a), the ratio between the scalar quark condensate and its value in the vacuum  $\langle \bar q q \rangle/\langle \bar q q \rangle_V$  (b), the light and strange quark effective masses $m^*_q, m^*_s$  (c), and the production probability of massive $s\bar s$ relative to light flavor production $\gamma_s$  (d) as a function of the energy density $\epsilon$ for the parameter sets NL3 (red solid lines), NL2 (thin orange lines)  and NL1 (dashed green lines) at $T=0$ and with $\Sigma_\pi$=45\, MeV.}
 \label{plots_EOS}
\end{figure}

In this section we analyze in more detail the flavor production ratios from the Schwinger formula in the presence of a hot and dense nuclear medium.
 As mentioned in the previous section  there are different sets for the parameters $g_s, m_\sigma, B, C$ in the gap equation (\ref{gap_equation}). In fact, these parameters are fixed within the non-linear $\sigma-\omega$ model in order to reproduce  empirical values  of nuclear matter quantities at saturation (i.e. saturation density, binding energy per nucleon, compression modulus, effective nucleon mass etc.), but sizeable uncertainties remain with respect to the high density properties.
In Table I we display the values of $g_s, m_s, B, C$ together with the vector coupling $g_v$, the vector meson mass $m_v$, the compression modulus $K$ and the ratio between the effective and the bare nucleon mass $m^*/m$ at saturation density for three sets commonly indicated as NL1, NL2 and NL3. The sets NL1 and NL3 have the same compression modulus $K$ but differ in the effective mass $m^*/m$ at saturation density whereas NL1 and NL2 have the same effective mass but differ in the compression modulus $K$. By comparing the results from NL1, NL2 and NL3 we will be able to explore separately the effects from the effective mass and compression modulus.
\begin{table}
\vspace{1em}
\label{tab1}
\begin{tabular}{|c|c|c|c|} \hline
                      & NL1    		& NL2  &  NL3  \\ \hline
 $g_s$                &  6.91  		&  8.50     & 9.50  \\ \hline
 $g_v$                &  7.54  		&  7.54    & 10.95  \\ \hline
 $B$ (1/fm)           &  -40.6 		&  50.57    & 1.589  \\ \hline
 $C$                  &   384.4 	&  -6.26   & 34.23  \\ \hline
 $m_s$ (1/fm)         &  2.79		&  2.79     & 2.79      \\ \hline
 $m_v$  (1/fm)        &  3.97		&  3.97     & 3.97      \\ \hline
 $K$ (MeV)            &  380		&  210      &  380       \\ \hline
 $m^*/m$              & 0.83		& 0.83     &   0.70       \\ \hline
\end{tabular}

\vspace{0.3cm} \caption{Parameter sets NL1, NL2 and NL3 for the
non-linear $\sigma-\omega$ model employed in the transport
calculations from Ref. \protect\cite{Lang}.}
\end{table}
In the context of the string decay, the most important parameter to
focus on is the scalar coupling $g_s$, which is lower for the NL1 and NL2 
set with respect to the corresponding values in the NL3  set. In
Fig. \ref{plots_EOS} we show the dependence of the nucleon scalar
density $\rho_S$ on energy density $\epsilon$ in panel (a),  the
ratio between the scalar quark condensate and its value in the
vacuum  $\langle \bar q q \rangle/\langle \bar q q \rangle_V$ in
panel (b), the light and strange quark effective masses $m^*_q,
m^*_s$ in panel (c), and the production probability of massive
$s\bar s$ relative to light flavor production $\gamma_s$ in panel
(d). Note that an analogous dependence is observed with respect to
the baryon density $\rho_B$ since the energy density $\epsilon$ in
leading order is just the nucleon mass times the baryon density. We find 
that all quantities plotted in Fig. \ref{plots_EOS} show practically identical results for NL1 (green dashed lines) and NL2 (thin  orange lines)
since  the scalar density $\rho_S$ essentially depends on the effective nucleon mass which is very similar for NL1 and NL2
when plotted as a function of the energy density $\epsilon$.
The results shown in Fig. \ref{plots_EOS} are obtained at vanishing
temperature $T=0$, but all the following considerations can be
extended to a more realistic picture at finite temperature (meson
density) and illustrate the consequences of CSR in the PHSD results
for heavy-ion collisions. The energy density $\epsilon$ here is
calculated within the non-linear $\sigma-\omega$ model by:
\begin{equation} \label{eps}
\epsilon = U(\sigma) + \frac{g_v^2}{2 m_v^2} \rho_N^2 + d \int
\frac{\mathrm{d}^3 p}{(2\pi)^3}
 \ E^*({\bf p}) \left(  N_{f}({\bf p})+  N_{\bar f}({\bf p})\right)\,, \end{equation}
with $$E^*({\bf p}) = \sqrt{{\bf p}^2 + m_N^{*2}}\,,$$
$$ U(\sigma) =
\frac{m_s^2}{2} \sigma^2 +  \frac{B}{3} \sigma^3 +  \frac{C}{4} \sigma^4\,, $$  where $\rho_N$
represents the nucleon density and $N_f({\bf p})$ and $N_{\bar f}({\bf p})$ are
the particle/antiparticle occupation numbers at fixed momentum ${\bf p}$, respectively.

The scalar density $\rho_S$ increases with increasing energy density
$\epsilon$ as displayed in panel (a) of Fig. \ref{plots_EOS}. We
find a moderate sensitivity to the nuclear equation of state up to
energy densities of $\sim$ 0.5 GeV/fm$^3$ that are of relevance for
the hadronic phase. In fact, the lines referring to the 
parameter sets NL1/NL2 and NL3 show a very similar behavior as a
function of $\epsilon$, but the NL3 (solid line) set is always
characterized by lower values of the scalar density $\rho_S$
relative to the NL1 or NL2 parametrization  (dashed line). This is due to
the larger value of the effective nucleon mass $m^*_N$ in case of the NL1 and NL2
parameter sets.  In panel (b) of
Fig. \ref{plots_EOS} the ratio $\langle \bar q q \rangle/\langle
\bar q q \rangle_V$ is presented as a function of $\epsilon$. At
$\epsilon=0$ the scalar condensate corresponds to the vacuum value
$\langle \bar q q \rangle_V$ and for fixed  $\Sigma_\pi$=45\, MeV it
decreases almost linearly with increasing energy density and almost
vanishes for the critical energy density $\epsilon_c \approx$ 0.5
GeV/fm$^3$. In this case, the order between NL1/NL2 and NL3 results is
reversed: the NL3 parametrization for the nuclear EoS is associated
to higher values of the scalar quark condensate with respect to the
NL1 or NL2 sets. This feature can be easily explained looking at the
definition of the ratio $\langle \bar q q \rangle/\langle \bar q q
\rangle_V$ (\ref{scalar_condensate}): at $T=0$, there are no thermal
mesons, thus the last term of the relation vanishes and the ratio is
entirely fixed by the scalar density $\rho_S$; hence, higher values
of $\rho_S$ correspond to lower values of $\langle \bar q q
\rangle/\langle \bar q q \rangle_V$. Therefore, the NL1 and NL2
parametrizations are characterized by lower values of the scalar quark
condensate with respect to the NL3 parameter set. It follows that
the light and strange quark effective masses $m^*_q, m^*_s$ in panel
(c) of Fig. \ref{plots_EOS} show a very similar dependence on the
energy density. At vanishing energy density $\epsilon$, the quark
effective masses are equal to their vacuum values,
$m_q\approx0.33\,$GeV and $m_s\approx0.5\,$GeV; with increasing
$\epsilon$ the quark masses decrease in line with the scalar quark
condensate up to their bare values $m_s^0 \approx 100\,$MeV and
$m_q^0 \approx 7\,$MeV for vanishing $\langle \bar q q
\rangle/\langle \bar q q \rangle_V$. The decrease of both $m_q$ and
$m_s$ is approximately linear in energy density  where the slope
associated to the light quark is flatter in comparison to the
strange quark mass. Concerning the comparison between the 
different choices for the nuclear equation of state, we find also
for these masses a non-negligible sensitivity and the same hierarchy
as for the scalar quark condensate (the results associated to NL1/NL2
are always below the corresponding results for NL3).

We recall that the effective masses of the quarks enter the
Schwinger formula (\ref{Schwinger-formula}) for the hadronic
particle production via the string decays. In panel (d) of Fig.
\ref{plots_EOS} the strangeness ratio $\gamma_s$ is shown as a
function of energy density for the two parameter sets. The factor
$\gamma_s$ increases from the vacuum case ($\gamma_s \approx$ 0.3)
with increasing energy density up to values of 0.8 -- 0.9 for
$\epsilon \approx \epsilon_c$. Thus the production of a $s\bar s $
pair relative to a light quark pair is no longer suppressed close to
the phase boundary for CSR as it is in vacuum. The reason of this
increase is the steeper decrease of the effective strange quark mass
(with energy density) in comparison to the effective light quark
mass as mentioned above. Furthermore, the NL1 and NL2 parametrizations give
larger values of $\gamma_s$ as the NL3 parametrization due to a
faster change of the masses with $\epsilon$ (cf. panel (c)).

\begin{figure}[t!]
\centering
\includegraphics[width=0.45\textwidth]{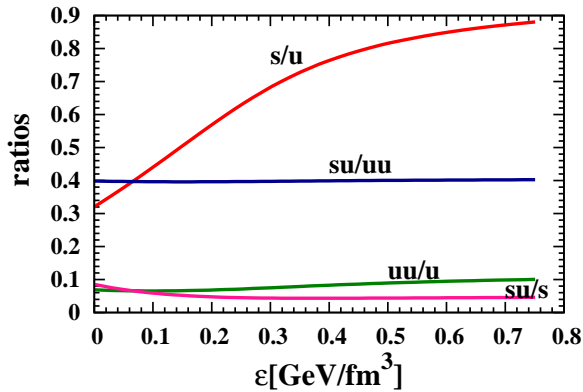}
\caption{(Color online) The quark and diquark ratios in the string decay (hadronic environment)
as a function of the energy density $\epsilon$
as evaluated within the non-linear $\sigma-\omega$ model
for the parameter set NL3 for $T=0$.}
 \label{fig_ratios}
\end{figure}

We note in extension of Ref. \cite{PHSD_CSR} that this scheme
for CSR in the string decay mechanism can be applied not only to the
light and strange quarks, but also to diquark combinations that are
additionally produced in the fragmentation of the string and lead
finally to baryon-antibaryon pairs. Here the default JETSET ratios
fix the diquark mass in the vacuum, e.g. a light diquark mass in
vacuum of $m_{uu}^V$= 0.65 GeV leads to a suppression of a light
diquark pair relative to a light quark-antiquark pair of
\begin{equation} \label{1} \frac{P(uu {\bar u} {\bar u})}{P(u {\bar
u})} \approx 0.07\,. \end{equation} For the creation of a diquark
($su$) one employs $m_{su}^V \approx$ 0.725 GeV which leads to the
ratio for a ($su$)-diquark pair relative to a light ($uu$)-diquark
pair of
\begin{equation} \label{2} \frac{P(su {\bar s}{\bar u})}{P(uu {\bar
u}{\bar u})} \approx 0.4\,. \end{equation} Within the same line the
vacuum mass of a $ss$-diquark can be determined from JETSET. The
Schwinger mechanism of string decay in vacuum thus requires the
following dressed vacuum masses: $ m_u^V \approx 0.35$ GeV, $m_s^V
\approx$ 0.5 GeV, $m_{uu}^V \approx$ 0.65 GeV, $m_{su}^V \approx$
0.725 GeV and $m_{ss}^V \approx$ 0.87 GeV to comply with
experimental observation in nucleon-nucleon collisions. The
production probability of diquarks ($su$) relative to $uu$-diquarks
(\ref{2}) and ($ss$) relative to $uu$-diquarks does not change very
much in the dense medium -- in line with (indirect) experimental
observation -- and since $m_{su}^0-m_{uu}^0 \approx
m_{s}^0-m_{u}^0$, the 'bare' diquark masses $m_{uu}^0$ can be fixed
and give $m_{uu}^0 \approx$ 0.5 GeV,  $m_{su}^0 \approx$ 0.593 GeV
and $m_{ss}^0 \approx$ 0.763 GeV. The explicit variations of the
flavor ratios with the energy density are displayed in Fig.
\ref{fig_ratios} and show that the diquark ratios only very
moderately change with the energy density whereas the $s/u$ ratio
steeply rises with $\epsilon$. We specify once more that these
results have been obtained within a pure hadronic system at
vanishing temperature, but the conclusions are valid also for finite
temperatures (meson densities).

The variations of the ratios in Fig. \ref{fig_ratios}, however, are
limited to the hadronic phase with energy densities below
$\epsilon_c$.
Above $\epsilon_c\approx0.5\,$GeV/fm$^{3}$ strings cannot be formed
anymore due to a vanishing string tension $\kappa$ in the QGP and
the hadrons dissolve to partonic degrees-of-freedom (and mean-field
energy). Consequently, the $s/u$ factor finally shows an increase
for $\epsilon < \epsilon_c$ (associated to CSR) and in
correspondence of $\epsilon \ge \epsilon_c$ it drops to the value
$\sim 1/3$ (fixed by comparison with the strangeness production at
RHIC and LHC energies observed experimentally). The energy
dependence of the $s/u$ ratio including the QGP phase has been shown
already in Fig. 2 of Ref. \cite{PHSD_CSR}. We recall that in the
partonic phase the $s/u$ ratio remains constant as a function of the
energy density. As a result we can identify a "horn" structure in
the $s/u$ ratio as a function of $\epsilon$, where the initial
increase is related to chiral symmetry restoration in the hadronic
phase and the subsequent sharp decrease is associated to the
formation of the QGP. We thus expect that CSR modifies the particle
abundances and spectra (especially in the strange particle sector)
from heavy-ion collisions, where increasing energy densities
$\epsilon$ in the overlap region can be achieved with increasing
bombarding energy (at the same centrality of the collision).

\subsection{Impact of $3M\leftrightarrow \mathrm{B}\mathrm{\bar B}$ collisions on the strangeness production}

For a solid study of the differential spectra of strange and
non-strange particles it is essential that both are treated on the
same level in many-body theory. Whereas in earlier PHSD studies the
three-body channels incorporating three mesons in the production of
baryon-antibaryon pairs in line with Ref. \cite{Cassing:2001ds}
(e.g. $\rho + \rho + \pi \leftrightarrow N + {\bar N}$) have been
incorporated by default in the non-strange sector  the corresponding
channels in the strangeness sector (e.g. $\rho+ K^* + \pi
\leftrightarrow {\bar N} + Y$ etc.) had been discarded in PHSD3.2.
Nevertheless, the possible impact of such channels on the
observables of interest in this study has to be examined in order to
control the consistency of the approach. Accordingly, in this
subsection we present the influence of an extended description of
$3\mathrm M\leftrightarrow \mathrm B\bar{\mathrm B}$ reactions to
the strange sector on the rapidity spectra from central heavy-ion
collisions with and without CSR.

\begin{figure}[t!]
    \centering
    \includegraphics[width=0.45\textwidth]{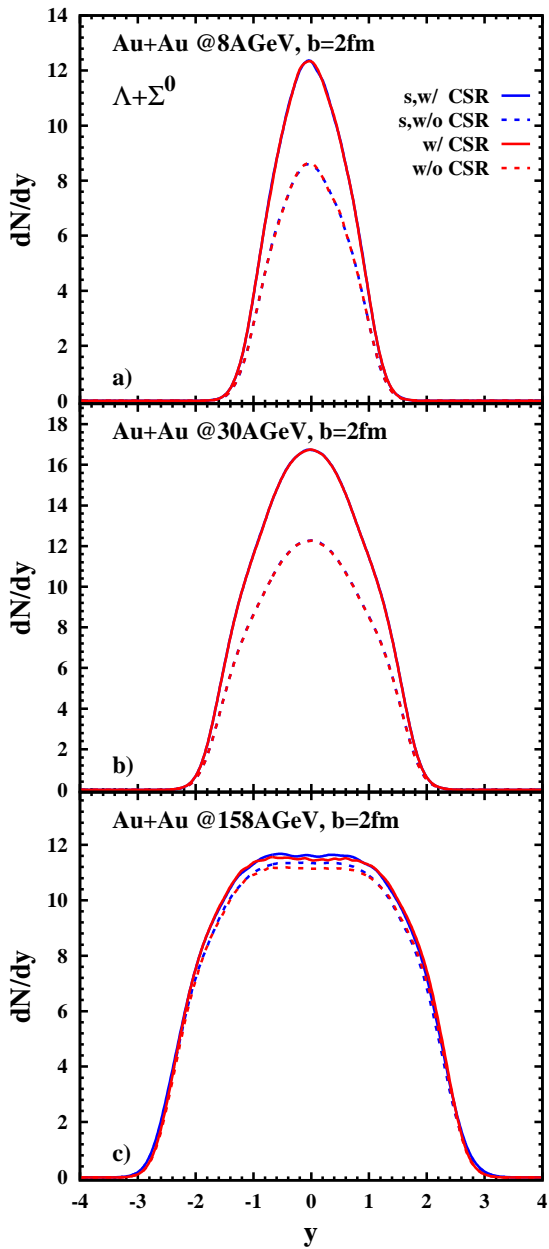}
    \caption{(Color online) Rapidity spectra for $\Lambda + \Sigma^0$ baryons in 5\% central
    Au+Au collisions for the energies of 8 (a), 30 (b) and 158\,AGeV (c) with (solid lines) and without CSR (dashed lines).
    The blue lines stand for the results that include the new
    three-meson reactions with strangeness content while the red
    lines display results when incorporating only non-strange
    three-meson channels.}
    \label{fig::old-new-3body}
\end{figure}

To this aim we have generalized the quark rearrangement model, first
presented in Ref. \cite{Cassing:2001ds} for the light ($u,d$)
sector, to the strange sector. The details of the extended model
will be given in an upcoming study where also the explicit tests of
detailed balance relations on a channel by channel basis are
demonstrated as well as the particular impact on multi-strange
antibaryon production and annihilation. By including the strangeness
sector the number of individual three-body (mass channels) amounts
to more than 2500 including also the fusion of three kaons (or
$K^*$'s) to the $N + {\bar \Omega}$ channel. We also consider the
hidden strangeness of the $\eta$ and the $\Phi$ mesons with proper
weights for channels where either the light quarks or $s\bar s$ are
rearranged. As in Ref. \cite{Cassing:2001ds} the matrix elements for
the transitions are provided by experimental data on
baryon-antibaryon annihilation.

Without going into further details we show in Fig.
\ref{fig::old-new-3body} -- exemplary for other particles -- the
rapidity spectra of $\Lambda+\Sigma^0$ baryons at 8, 30 and
158\,AGeV in central Au + Au collisions with (solid lines) and
without (dashed lines) chiral symmetry restoration as well as with
(blue lines)  and without (red lines) the extended strangeness
$3\mathrm{M}\leftrightarrow\mathrm{B}\mathrm{\bar{B}}$ reactions. In
general, the inclusion of chiral symmetry restoration enhances the
number of produced $\Lambda+\Sigma^0$ baryons at all energies. This
is most pronounced  for the lower energies of 8 and 30 AGeV (cf. the
detailed studies in the next section). In particular for the lower
energies there is no visible difference in the rapidity spectra for
the extended $3\mathrm{M}\leftrightarrow\mathrm{B}\bar{\mathrm{B}}$
reactions. Only at 158\,AGeV one finds a slight difference ($\sim$ 1
\%) in the central rapidity region. The reason for these findings is
related to the low abundance of mesons carrying a strange quark at
low bombarding energies and additionally in the heavier masses of
the hyperons which induce a lower phase-space for production, since
the transition probability is directly proportional to the available
phase-space in the flavor rearrangement model of Ref.
\cite{Cassing:2001ds}. We conclude that the extended three-body
reactions with strangeness have no crucial impact at AGS and SPS
energies.

\section{Application to nucleus-nucleus collisions}
In this section we study observables from heavy-ion collisions
with respect to different novel aspects that had not been considered
in Ref. \cite{PHSD_CSR}. In this respect we present results  for the
rapidity distribution of the most abundant particles at AGS and SPS
energies for different nuclear equations of state in order to
estimate the uncertainties of our approach. In addition we also
explore the impact of CSR on the transverse dynamics by calculating
the transverse mass spectra for protons, pions, kaons and antikaons
in comparison to available data. We recall that in particular the
transverse slopes of the kaon spectra had been clearly
underestimated in the earlier HSD studies (without a partonic phase)
\cite{Brat04}. As a survey we present the excitation functions of
particle yields and ratios together with the uncertainty due to the
nuclear EoS. Moreover, we show the time evolution of the number of
strange particles and the corresponding production rates from the
different production channels. The following scenarios will be
explored:
\begin{itemize}
\item default PHSD calculations without CSR;
\item PHSD calculations including CSR with NL3 as parameter set for the nuclear EoS;
\item PHSD calculations including CSR with NL2 as parameter set for the nuclear EoS;
\item PHSD calculations including CSR with NL1 as parameter set for the nuclear EoS.
\end{itemize}

\subsection{Time evolution of the strange particle multiplicities}
\begin{figure}[h!]
\centering
\includegraphics[width=0.45\textwidth]{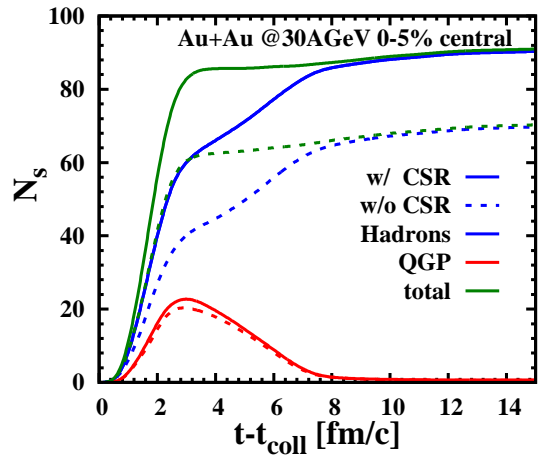}
\caption{(Color online) The strange particle number $N_s$ as a function of time in 5\% central Au+Au collision at 30 \,AGeV.
The solid lines show the results from PHSD including CSR with NL1 parameters, while the dashed lines show the results from PHSD without CSR. The green lines refer to the total number of strange particles, while the blue and red lines refer to the hadronic and partonic contributions of the strangeness content, respectively.}
 \label{number_strange}
\end{figure}

At AGS and SPS energies the strange particle production takes place
at the early stages of the collision process as it is seen from Fig.
\ref{number_strange}, where the number of particles containing
$s$-quarks $N_s$ is plotted (green solid and dashed lines) as a
function of time in central Au+Au collision at a bombarding energy
of 30 \,AGeV for 5\% most central collisions.
More than $\approx90\%$ of the strange content of the system is
created in the time interval between 0 and 4\,fm/c after the
collision and already at about 8\,fm/c strange particles are not
produced anymore. In Fig. \ref{number_strange} the total number of
strange particles has also been separated in the corresponding
hadronic and partonic contributions, represented by the blue and red
lines, respectively. At the energy of 30\,AGeV, the hadronic
strangeness content is dominant relative to the partonic one, which
can reach  $\sim 25\%$ of the total strangeness content in
correspondence of the maximum value of the partonic contribution.
The strange quarks in the partonic phase appear in the system not
immediately after the collision, but after about 1.5\,fm/c. In fact,
the primary interactions within the PHSD are realized via string
excitation and after that in the cells with energy density
$\epsilon>0.5\,$GeV/fm$^{3}$ the hadrons are dissolved into partons
and mean field energy. The partonic $N_s$ distribution initially
increases as a function of time, reaching a maximum at about
3\,fm/c. Then at larger times the energy density of the system
decreases, as a result the partons hadronize by dynamical
coalescence which lasts up to about 8\,fm/c, when the strange hadron
number is basically fixed. We can conclude that at 30\,AGeV the
steep increase of the number of strange particles as a function of
time has to be attributed dominantly to the hadronic production,
which occurs in PHSD via string formation and decay. The other
hadronic scattering processes, which are not negligible at this
bombarding energy, do not create further strangeness in the system,
but they are only responsible for strange flavor exchanges.

It is, furthermore, interesting to compare the strange particle
amount computed in PHSD including and excluding CSR in the string
dynamics. In Fig. \ref{number_strange} we show PHSD calculations
only in the limits: without CSR (dashed lines) and including CSR
with NL1 as parameter set (solid lines). The restoration of chiral
symmetry causes a sizable increase ($\approx30\%$) of the total
strangeness content. We notice that CSR does not modify the time
evolution of $N_s(t)$, but it only affects the hadronic contribution
to the strange particle production. There is a slight difference
between the partonic results with and without CSR since the strange
particle number in the partonic phase is slightly higher when
including CSR. This is, however, not due to a higher strangeness
production in the QGP, but stems from particles, which are produced
by string decay in the hadronic corona and travel to cells with
energy density above $\epsilon_c$  during their propagation. Thus
such strange particles, even if produced by the string decay,
dissolve into partonic degrees-of-freedom. In this respect the
enhancement of the strange particle number in the hadronic phase
drives a small increase also in the partonic contribution of $N_s$.

\begin{figure}[t!]
\centering
\includegraphics[width=0.45\textwidth]{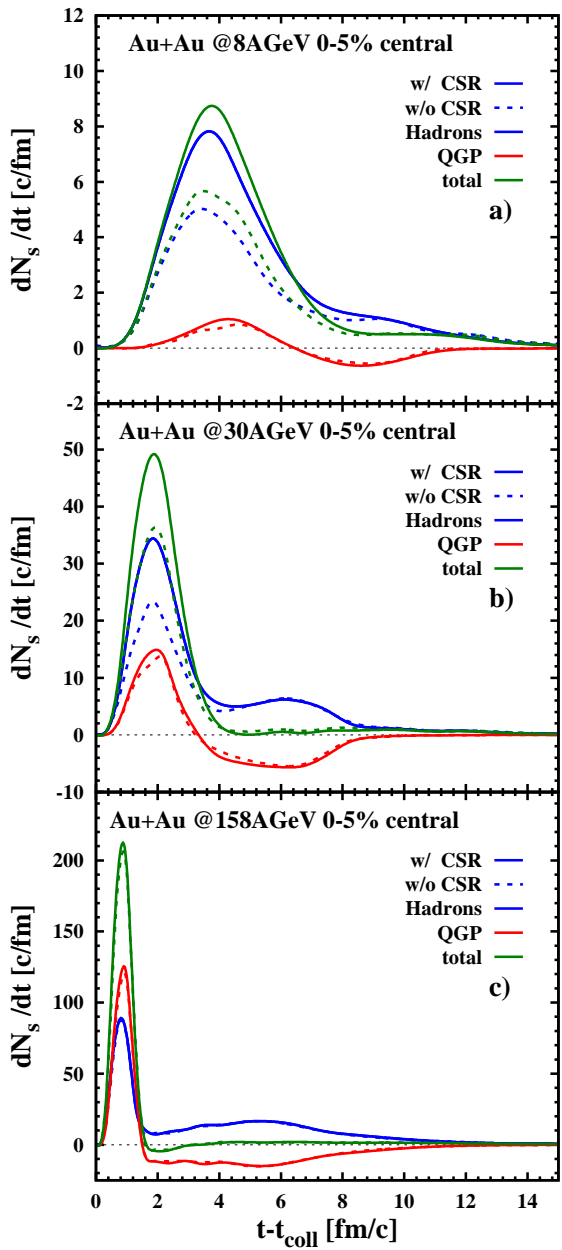}
\caption{(Color online) The strangeness rate $dN_s/dt$ as a function
of the time in 5\% central Au+Au collisions at 8 (a), 30 (b),
158\,AGeV (c). The coding of the lines is the same as in Fig.
\ref{number_strange}. Note that these rates become negative when
strange hadron dissolution or strange parton hadronization dominate,
respectively. }
 \label{rate_strange}
\end{figure}

To provide a more complete illustration of the time evolution of the
strangeness content in a heavy-ion collision we show in Fig.
\ref{rate_strange} the rate $dN_s/dt$ of  strange particles  at 8
(a), 30 (b), 158\,AGeV (c) in central collisions. We show again the
total strange particle rate in green, the hadronic contribution in
blue and the partonic contribution in red, comparing in all cases
the calculations with and without CSR by the solid and dashed lines,
respectively. We mention that these rates include production as well
as losses either due to dissolution of hadrons in the QGP or due to
hadronization of strange partons. Accordingly these rates become
negative when dissolution or hadronization dominates.
As already seen in Fig. \ref{number_strange}, the strange particle
production occurs at the first stages of the collisions and the
strangeness production rate increases with increasing bombarding
energy. The hadronic contribution is dominant at the collision
energies $E_{Lab}=8$ and 30\,AGeV, while the partonic production
rate is larger than the hadronic one at $E_{Lab}=158$\,AGeV. The
strangeness enhancement associated to CSR is most clearly visible at
lower energies while  at 158\,AGeV it is very moderate. We will find
the same result on the final particle rapidity spectra (see next
subsection). The total strange particle rate remains positive during
the entire time evolution in panels (a) and (b) and shows a small
negative rate only for $E_{Lab}=158$\,AGeV due to the dissolution of
strange hadrons in the QGP; on the other hand the partonic rate
becomes negative at larger times since the hadronization ('loss')
dominates the strange quark production. The negative rate on the
partonic side is balanced by the positive rate on the hadronic side
due to strangeness conservation.

Comparing the rates at the different energies, we can see that the
strangeness production is slower at lower energies. In fact, the
peak of the total $dN_s/dt$ shifts to smaller times with increasing
energies, and for $E_{Lab}=158\,$AGeV the whole strangeness
production occurs within $2\,$fm/c. On the other hand the duration
of the hadronization process becomes longer at higher bombarding
energies, where a larger volume of the systems turns into the QGP
phase. In general, after $\approx12\,$fm/c the strangeness content
is fixed and within PHSD the creation of strange particles ceases.
Equilibrium aspects of the strangeness production will be discussed
in Sec.VI.

\subsection{Rapidity spectra at AGS and SPS energies}
\begin{figure}[t!]
\centering
\includegraphics[width=0.43\textwidth]{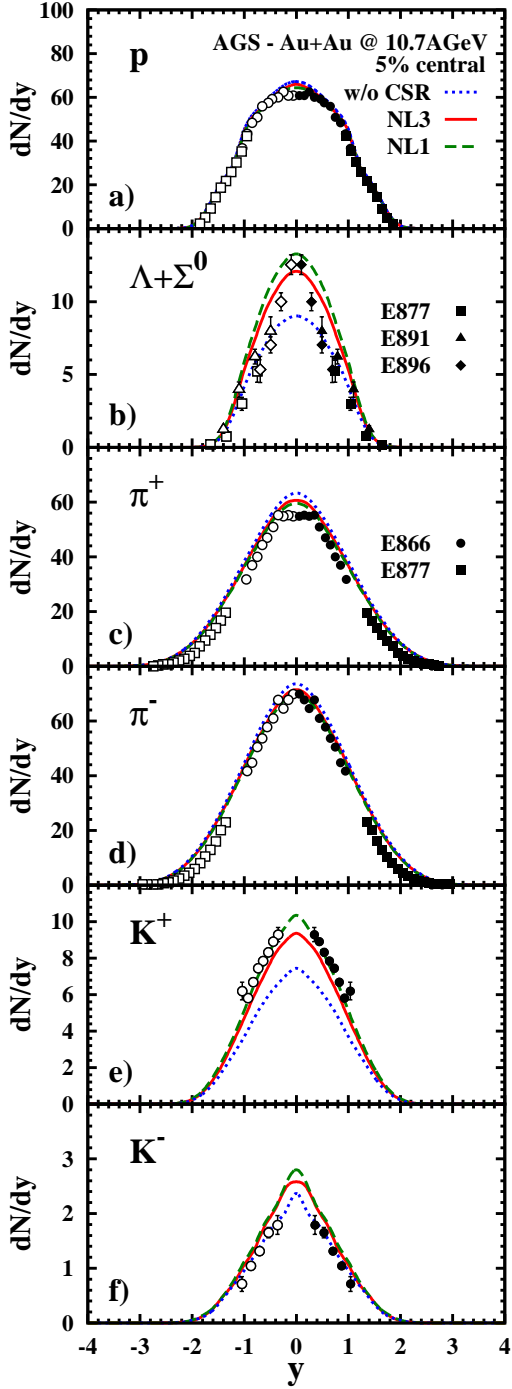}
\vspace{-0.75cm}\caption{(Color online) The rapidity distribution of protons, $(\Lambda+\Sigma^0)$'s, pions and kaons for 5\% central Au+Au collisions at 10.7\,AGeV in
comparison to the experimental data from Refs. \cite{exp1,exp1b}. The solid
(red) lines show the results from PHSD including CSR with NL3 parameters, the dashed green lines show the results from PHSD including CSR with NL1 parameters and the blue dotted lines show the result from PHSD without CSR.}
 \label{rapidity1}
\end{figure}

\begin{figure}[t!]
\centering
\includegraphics[width=0.43\textwidth]{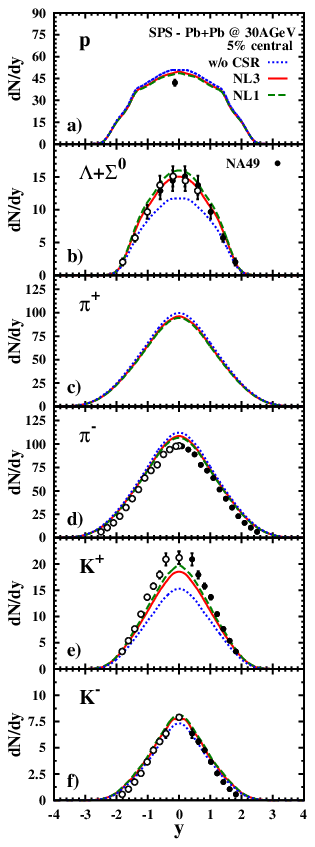}
\vspace{-0.75cm}\caption{(Color online) The rapidity distribution of protons, $(\Lambda+\Sigma^0)$'s, pions and kaons
for 5\% central Au+Au collisions at 30\,AGeV in comparison to the experimental data from Ref. \cite{exp2a,exp2b,exp2c}.
The coding of the lines is the same as in Fig. \ref{rapidity1}.}
 \label{rapidity2}
\end{figure}

\begin{figure}[t!]
\centering
\includegraphics[width=0.43\textwidth]{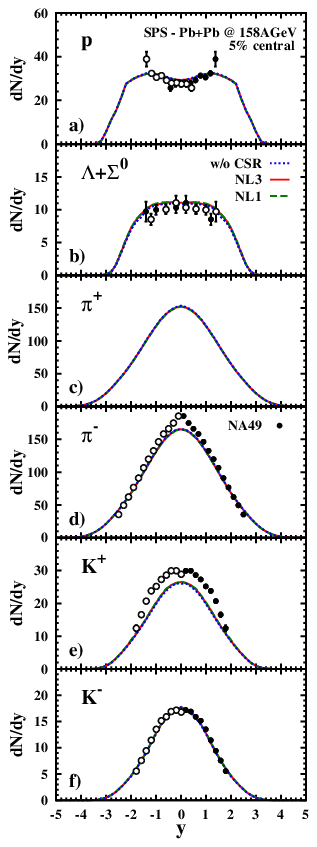}
\vspace{-0.75cm}\caption{(Color online) The rapidity distribution of
protons, $(\Lambda+\Sigma^0)$'s, pions and kaons for 5\% central
Au+Au collisions at 158A\,GeV in comparison to the experimental data
from Ref. \cite{exp3a,exp3b}. The coding of the lines is the same as in
Fig. \ref{rapidity1}.}
 \label{rapidity3}
\end{figure}

We present in Figs. \ref{rapidity1} to \ref{rapidity3} the PHSD
results for the rapidity distribution of protons,
$(\Lambda+\Sigma^0)$'s, pions and kaons for central nucleus-nucleus
collisions at different energies (from AGS to top SPS energies) in
comparison to the experimental data from Refs.
\cite{exp1,exp1b,exp2a,exp2b,exp2c,exp3a,exp3b}. The following scenarios will be
explored at different energies:
\begin{itemize}
{\item default PHSD calculations without CSR, represented by the dotted blue lines;}
{\item PHSD calculations including CSR with NL3 as parameter set for the nuclear EoS, represented by the solid red lines;}
{\item PHSD calculations including CSR with NL1 as parameter set for the nuclear EoS, represented by the dashed green lines.}
\end{itemize}
We note that PHSD calculations for the parameter set NL2 are not shown explicitly
since the results are in between those for NL1 and NL3. We will come back to an explicit comparison below. 
The results of the first two scenarios are almost equivalent to the
PHSD calculations shown in the previous study \cite{PHSD_CSR}, with
a slight difference regarding the proton and meson spectra. This is
due to a few recent upgrades in PHSD: The first concerns the
extension of inelastic meson-meson scattering above 1.3 GeV of
invariant energy $\sqrt{s}$ to string formation and decay with a
cross section of $\sim$ 10 mb; the second improvement is related to
p-wave scattering in the reaction channel $\pi + N \leftrightarrow
\Delta$  which had been treated isotropically before. Note,
however, that the study in Ref. \cite{PHSD_CSR} had a focus on the
comparison of HSD and PHSD calculations in order to examine also the
role of partonic degrees-of-freedom whereas here we concentrate on
the variation of the spectra with respect to the nuclear EoS
employing only the updated PHSD3.3 calculations.

First, we compare the results with and without CSR at
$E_{Lab}=10.7\,$AGeV (Fig.  \ref{rapidity1}), to point out the
general effect of this mechanism on the final particle rapidity
distributions and in particular show the dependence on the
parameter sets NL1 and NL3 for the non-linear $\sigma-\omega$ model
for the nuclear EoS in extension to Ref. \cite{PHSD_CSR}. The
restoration of chiral symmetry gives an enhancement of the strange
particle yields both for mesons and baryons. On the other hand, it
produces a slight decrease in the number of pions at midrapidity due
to the suppression of pions in the string decays in favor of strange
hadrons. The proton rapidity spectra do not present any sensible
variation, in fact the CSR as implemented in PHSD modifies
essentially the chemistry of the newly produced particles in the
string decay and has a minor impact the dynamics of the nucleons,
which in the string picture are associated to the string ends of the
primary interactions in the system. The inclusion of the CSR is
essential in order to correctly reproduce the strange particle
rapidity spectra, as we can see especially for $(\Lambda+\Sigma^0)$
hyperons and $K^+$ mesons. Furthermore, our calculations for the
proton rapidity spectra are in good agreement with experimental
observation.

Next, we discuss the results from PHSD with CSR using two different
parametrization for the nuclear equation of state, i.e. NL3 and NL1.
The general features of the strangeness enhancement hold for both
parametrizations; in particular the NL1 set provides larger values
for all strange particle rapidity spectra at midrapidity in line
with the discussion of Fig. 1. The difference between the two
parametrizations represents the uncertainty of our results related
to CSR as implemented in PHSD. We stress that we do not tune the
parameters of the equation of state to fit the data, but we employ
different nuclear EoS  to compute the scalar density (as explained
in section \ref{EOS}) in order to explore the uncertainties of our
approach.

In Fig. \ref{rapidity2} the rapidity spectra of various hadrons at
$E_{Lab}=30\,$AGeV are shown. We find the same features as for
$E_{Lab}=10.7\,$AGeV concerning the strangeness enhancement and the
comparison between the two parameter sets for the equation of state;
the differences are slightly smaller at this energy. At midrapidity,
both protons and pions are very slightly over-estimated in all
explored scenarios which suggests that the nuclear stopping is still
a bit overestimated. Finally, at the top SPS energy
$E_{Lab}=158\,$AGeV (Fig. \ref{rapidity3}) the CSR does not play a
significant role, since the dynamics is dominated by the QGP phase.
Thus, there is no appreciable difference between the results with
and without CSR for the two different EoS. Our results for $\pi^-$
and $K^+$ are lower with respect to the experimental data, however,
the $(\Lambda+\Sigma^0)$ and $K^-$ as well as the protons are
correctly reproduced. It is presently unclear where these final
differences stem from, since strangeness conservation is exactly
fulfilled in the PHSD calculations.

As we have seen in Fig. 1 the results for the strangeness ratio $\gamma_s(\epsilon)$ 
are very similar for the parameter sets NL1 and NL2 for nuclear matter at $T$=0.
In heavy-ion collisions, however, the different compression modulus leads to 
a slightly different baryon dynamics which also has an impact on the meson abundances and spectra.
In order to quantify the effect of the different nuclear EoS on the particle abundances
we provide in table II the midrapidity densities for protons, pions, $K^+, K^-$ and $\Lambda + \Sigma^0$
as well as their ratios for the parameter set NL1, NL2 and NL3 in case of a central Pb+Pb 
collision at  30\,AGeV, where the effect from CSR is most pronounced. As one can extract from the table, 
the proton and pion densities at midrapidity are correlated: a higher stopping goes along with a higher 
pion density with the order NL3 $>$ NL2 $>$ NL1. On the other hand the strangeness densities at midrapidity
are anticorrelated with the proton density; we obtain the order NL1 $>$ NL2 $>$ NL3. Although the hadron
densities differ not so dramatic it gives an enhanced effect in ratios $K^+/\pi^+$, $K^-/\pi^-$ and $(\Lambda + \Sigma^0)/\pi$
in the order NL1 $>$ NL2 $>$ NL3. Since any realistic nuclear EoS is expected to provide results within the
limits of these parameter sets we expect to obtain reliable bounds on the uncertainties with respect to the nuclear EoS.

\begin{table}
\vspace{1em}
\label{tab2}
\begin{tabular}{|c|c|c|c|} \hline
                      & NL1    		& NL2  &  NL3  \\ \hline
 p                &  47.6 		&  48.2     & 48.5  \\ \hline
 $\pi^+$                & 91.1  	&  91.7    & 92.5  \\ \hline
 $\pi^-$           &  102.7 		&  103.3    & 104.2  \\ \hline
 $K^+$                  &   18.6 	&  18.1   & 17.6  \\ \hline
 $K^-$                  &   7.58 	&  7.45   & 7.34  \\ \hline
 $\Lambda+\Sigma^0$              & 15.6	& 15.1     &  14.7       \\ \hline
 $K^+/\pi^+$                  &  0.204 	&  0.197   & 0.190  \\ \hline
 $K^-/\pi^-$                  &  0.0738 	&  0.0721   & 0.0704  \\ \hline
 $(\Lambda+\Sigma^0)/\pi$              & 0.0537		& 0.0516     &   0.0498       \\ \hline
\end{tabular}
\vspace{0.3cm} \caption{Particle abundances and strange to non-strange particle ratios at midrapidity ($|y|\le0.5$) from 5\% central Pb+Pb collisions at 30\,AGeV.}
\end{table}

\subsection{Transverse mass spectra at AGS and SPS energies}
\begin{figure}[t!]
\centering
\includegraphics[width=0.45\textwidth]{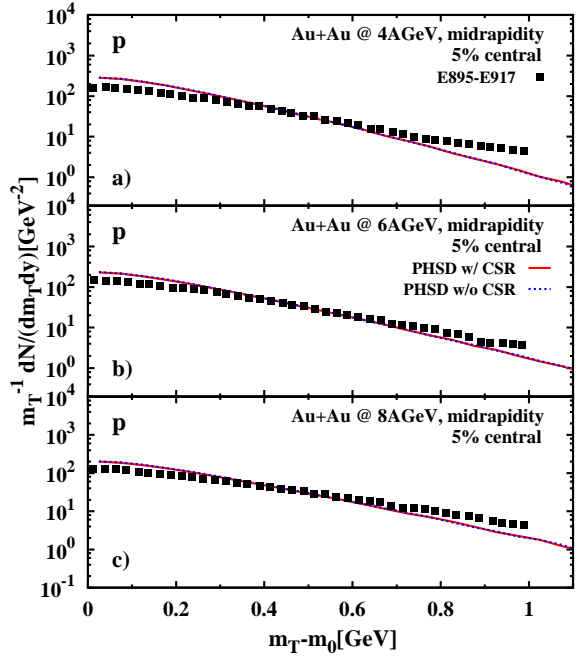}
\caption{(Color online) The transverse mass spectra of protons for
5\% central Au+Au collisions at $4,6,8\,$AGeV in comparison to the
experimental data from Ref. \cite{expmt1}. The solid (red) lines
show the results from PHSD including CSR with NL3 parameters, the
dotted (blue) lines show the results from PHSD without CSR.}
 \label{mtp1}
\end{figure}

\begin{figure}[h!]
\centering
\includegraphics[width=0.48\textwidth]{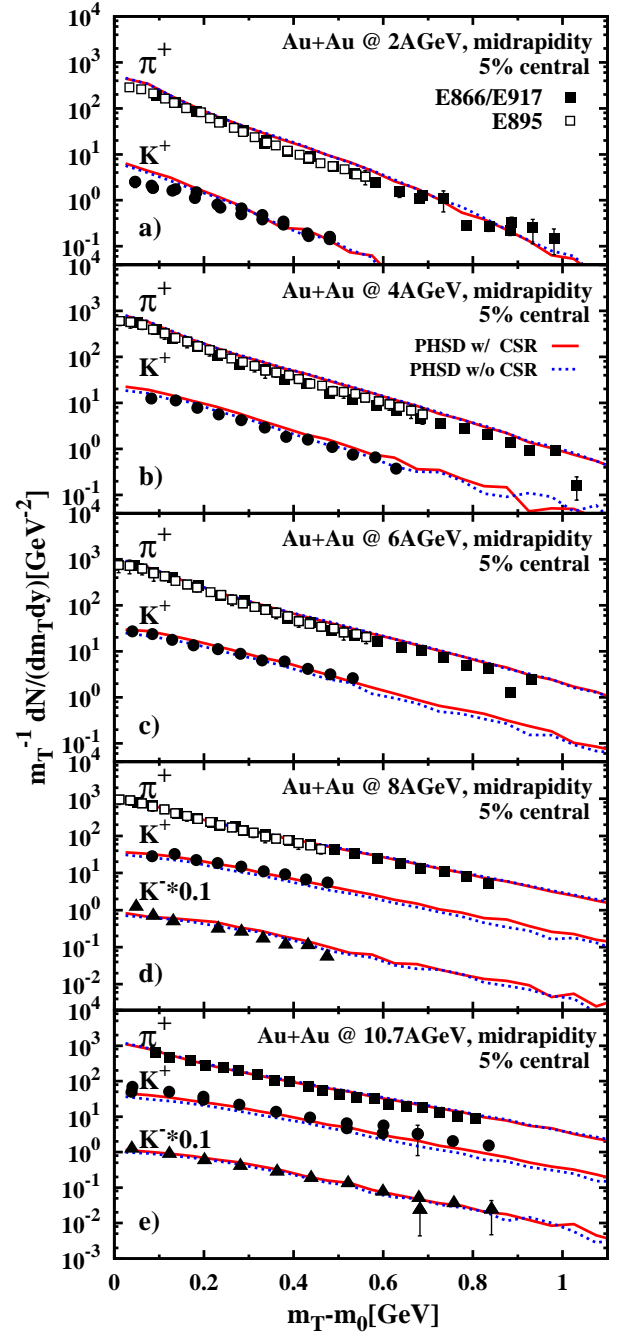}
\vspace{-0.75cm}\caption{(Color online) The transverse mass spectra of pions and kaons
for 5\% central Au+Au collisions at $2,4,6,8,10.7\,$AGeV in comparison to the experimental data from Refs. \cite{expmt3a,expmt3b}.
We show the results from PHSD including CSR with NL3 parameters by solid (red) lines and those from PHSD without CSR by dotted (blue) lines.}
 \label{mtm1}
\end{figure}

\begin{figure}[h!]
\centering
\includegraphics[width=0.45\textwidth]{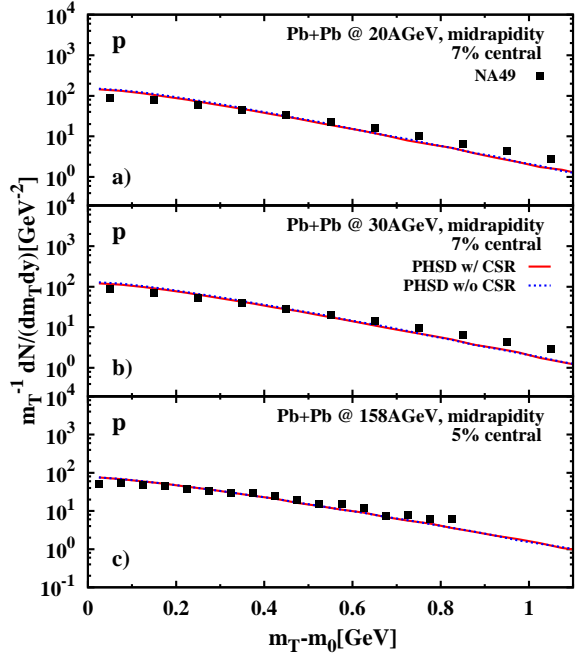}
\caption{(Color online) The transverse mass spectra of protons
for 5\% and 7\% central Pb+Pb collisions at $20,30,158\,$AGeV in comparison to the experimental data from Ref. \cite{expmt2}.
The coding of the lines is the same as in Fig. \ref{mtp1}.}
 \label{mtp2}
\end{figure}

\begin{figure}[h!]
\centering
\includegraphics[width=0.48\textwidth]{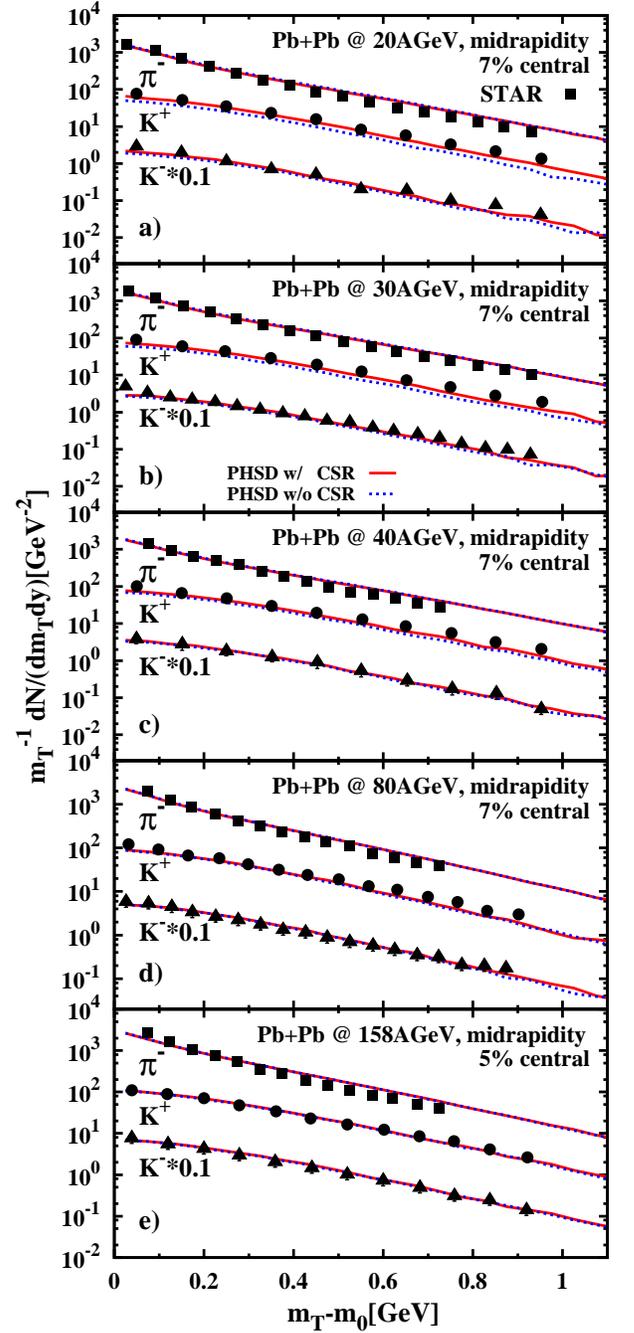}
\vspace{-0.75cm}\caption{(Color online) The transverse mass spectra of pions and kaons
for 5\% and 7\% central Pb+Pb collisions at $20,30,40,80,158\,$AGeV in comparison to the experimental data from Ref. \cite{exp2a,exp3a,expmt4}.
The coding of the lines is the same as in Fig. \ref{mtm1}.}
 \label{mtm2}
\end{figure}

We recall that in earlier HSD calculations (without a partonic
phase) the slopes of the transverse mass distributions have been
severely underestimated \cite{Brat04}. In extension to Ref.
\cite{PHSD_CSR} we show in this section the PHSD results for the
transverse mass spectra of protons, pions and kaons for different
energies in central Au+Au collisions in Figs. \ref{mtp1} and
\ref{mtm1} and Pb+Pb collisions in Figs. \ref{mtp2} and \ref{mtm2}
in comparison with AGS and SPS data, respectively. We do not show
the further scenarios (PHSD calculations including CSR with NL1 or NL2) since
the differences with respect to the parameter set NL3 for CSR is
practically not visible.

At AGS energies (Fig. \ref{mtp1}), i.e.  $E_{Lab}=4,6,8\,$AGeV, our
calculations for the proton spectra show the same trend as the
experimental data. However, we observe that the computed spectra are
softer than the experimental data in this energy regime. In fact,
our results over-estimate the data at low transverse mass $m_T$  and
under-estimate the data at larger $m_T$. Note, however, that in our
present calculations hadronic potentials have not been included. We
leave the study of explicit hadronic potentials in the propagation
of the degrees-of-freedom to a future work. We notice that CSR
produces no change in the transverse mass spectra of the protons,
both at AGS energies (Fig. \ref{mtp1}) and at SPS energies (Fig.
\ref{mtp2}). At larger energies, i.e. $E_{Lab}=20,30,158\,$AGeV
(Fig. \ref{mtp2}), the PHSD results are in a good agreement with the
experimental data for protons. In the latter cases, a sizeable
volume of the system performs a phase transition to the QGP and the
final particle spectra are not sensitive to hadronic potentials
anymore because the baryon densities in the final hadronic phase are
rather low.

In Figs. \ref{mtm1} and \ref{mtm2} we display the transverse mass
spectra for pions and kaons in central Au+Au collisions at AGS
energies and in Pb+Pb collisions at SPS energies, respectively. We
focus on the role played by the CSR on the mesons transverse mass
spectra. At the lower energies, $E_{Lab}=2\,$AGeV there is no
appreciable difference between the calculation with and without CSR,
since the energy density reached by the system is not high enough to
produce a vanishing scalar quark condensate. Instead, in the energy
range $E_{Lab}=4-40\,$AGeV, we notice a small difference between the
scenarios with and without CSR. As already mentioned, the CSR acts
directly on the chemistry and not so much on the dynamics of the
Schwinger mechanism, thus the effect of the partial restoration of
chiral symmetry is rather small on the transverse mass spectra. The
kaon spectra are harder when CSR is included, while the pion spectra
remain essentially unchanged. At the higher SPS energies
$E_{Lab}=80,158\,$AGeV the dynamics of the system is ruled
dominantly by the QGP phase and our calculations do not show any
sensitivity on the inclusion of CSR. The agreement of our PHSD
calculations with the data in Figs. \ref{mtm1} and \ref{mtm2} is
good in all cases studied. Even if the focus of this work is the
study of CSR, we point out that additional kaon potentials might
modify this picture at low energies. In particular, the attractive
potential for $K^-$ in the hadronic phase should improve our
calculations at $E_{Lab}=8\,$AGeV producing a softening of the
spectra. We will report on the effect of hadronic potentials in a
forthcoming study.

\subsection{Strange particle abundances and ratios}
\begin{figure}[t!]
\centering
\includegraphics[width=0.48\textwidth]{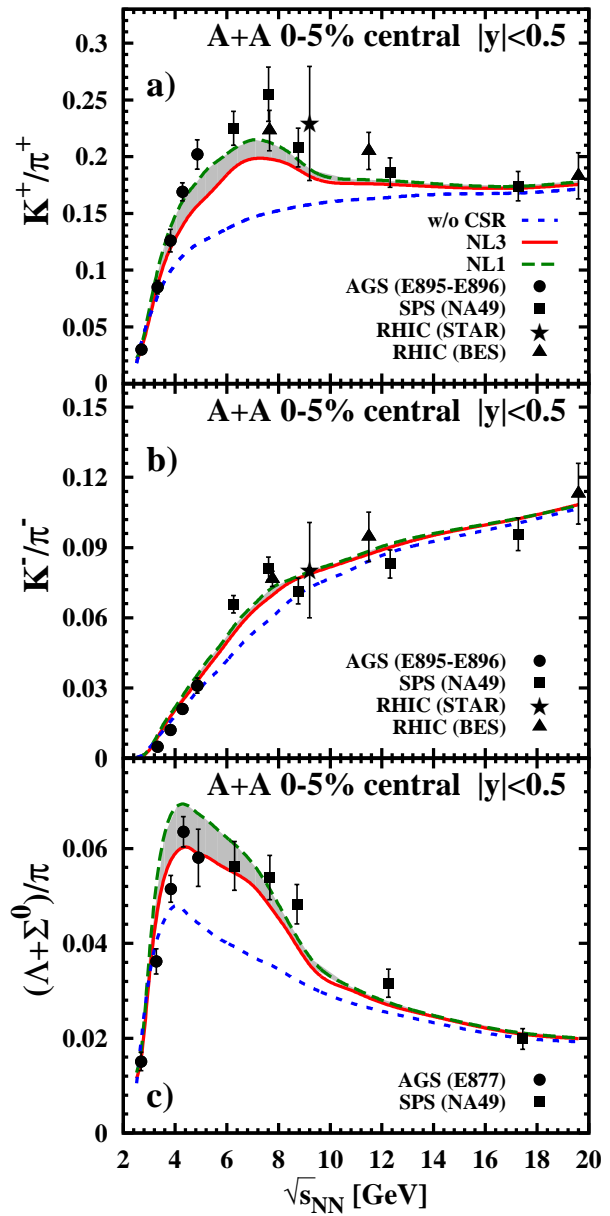}
\vspace{-0.7cm}\caption{(Color online) The ratios $K^+/\pi^+$,
$K^-/\pi^-$ and $(\Lambda+\Sigma^0)/\pi$ at midrapidity from 5\%
central Au+Au collisions as a function of the invariant energy
$\sqrt{s_{NN}}$ up to the top SPS energy in comparison to the
experimental data from \cite{exp4,expmt3b,exp2c}. The coding of the
lines is the same as in Fig. \ref{rapidity3}. The grey shaded area
represents the results from PHSD including CSR taking into account
the uncertainty from the parameters of the $\sigma-\omega$-model for
the EoS.}
 \label{horn}
\end{figure}

In this subsection we study the excitation function of the particle
ratios $K^+/\pi^+$, $K^-/\pi^-$ and $(\Lambda+\Sigma^0)/\pi$ at
midrapidity from 5\% central Au+Au collisions. In Fig. \ref{horn} we
show the calculations for the following three scenarios: the default
PHSD without CSR (blue dotted line), PHSD including CSR with NL3 and
NL1 as parameter sets for the nuclear EoS from the non-linear
$\sigma-\omega$ model (red solid and green dashed lines,
respectively). The shaded area displays the uncertainties of our
calculations from the two scenarios for the nuclear EoS since the results 
from the parameter set NL2 are always in between those from NL1 and NL3 (cf. table II).
As already described in Ref. \cite{PHSD_CSR}, the inclusion of CSR
in PHSD is responsible for the strong strangeness enhancement at AGS
and low SPS energies. The experimental observations of the ratios
$K^+/\pi^+$ and $(\Lambda+\Sigma^0)/\pi$ show the well-known "horn"
structure, which is reproduced by the PHSD calculations with CSR. In
fact, CSR gives rise to a steep increase of these ratios at energies
lower than $\sqrt{s_{NN}}\approx 7\,$GeV, while the drop at larger
energies is associated to the appearance of a deconfined partonic
medium. As anticipated by the considerations in Sec. \ref{EOS}, the
NL1 parameter set produces a sharper peak both in the $K^+/\pi^+$
and in the $(\Lambda+\Sigma^0)/\pi$ excitation functions with a
$\approx 10\%$ maximum increase with respect to the NL3 result
that had been reported in Ref. \cite{PHSD_CSR}. We point out that
even adopting different parametrizations for the $\sigma-\omega$
model, we recover the same "horn" feature. This supports the
reliability of the CSR mechanism as implemented in the PHSD model.

At AGS energies, the energy dependencies of the ratios $K^+/\pi^+$
and $(\Lambda+\Sigma^0)/\pi$ are closely connected, since $K^+$ and
$\Lambda$ (or $\Sigma^0$) are mostly produced in pairs due to
strangeness conservation. On the other hand, the excitation function
of the $K^-/\pi^-$ ratio does not show any peak, but it smoothly
increases as a function of $\sqrt{s_{NN}}$. In fact, especially at
AGS energies, the antikaon production differs substantially from the
production of $K^+$ and $\Lambda$, which occurs dominantly via
string formation. In fact, the antikaons are produced mainly via
secondary meson-baryon interactions by flavor exchange and their
production is suppressed with respect to the $\Lambda$ hyperons that
carry most of the strange quarks. This is the reason why the
inclusion of chiral symmetry restoration provides a substantial
enhancement of the $K^+/\pi^+$ and $(\Lambda+\Sigma^0)/\pi$
excitation functions and a smaller change on the $K^-/\pi^-$ ratio.
We also notice that there is no  sizeable difference between the NL1
and NL3 results for the $K^-/\pi^-$ ratio. At top SPS energies the
strangeness is produced predominantly by the hadronization of
partonic degrees-of-freedom, thus our results for all the ratios do
not show an appreciable sensitivity to the nuclear EoS and the
calculations with and without CSR tend to merge at
$\sqrt{s_{NN}}\approx20\,$GeV.

\begin{figure}[t!]
\centering
\includegraphics[width=0.48\textwidth]{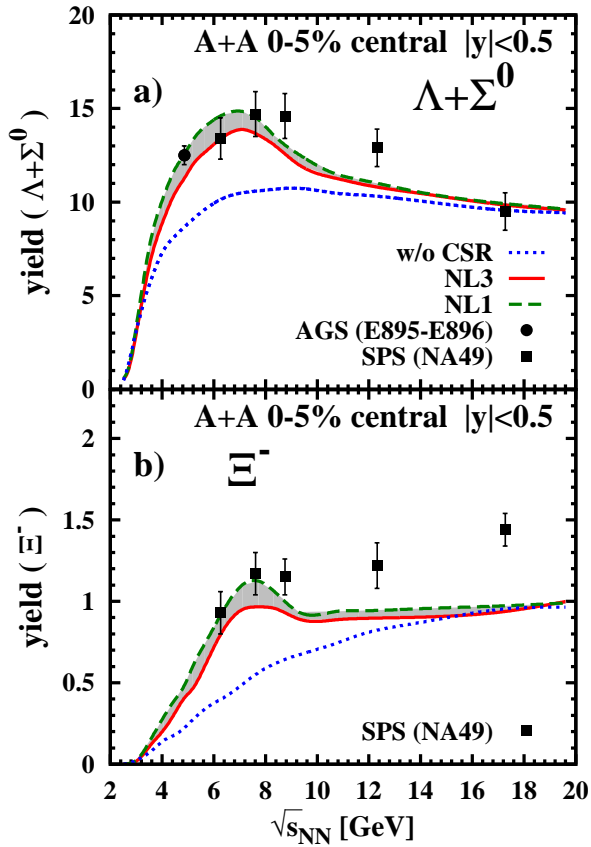}
\vspace{-0.7cm}\caption{(Color online) The yields of $(\Lambda+\Sigma^0)$ and $\Xi^-$ at midrapidity from 5\% central Au+Au collisions as a function of the invariant energy $\sqrt{s_{NN}}$ up to the top SPS energy in comparison to the experimental data from Refs. \cite{exp1b,exp2c}. The coding of the lines is the same as in Fig. \ref{horn}.}
 \label{hyperons2}
\end{figure}

Finally, in Fig. \ref{hyperons2} we present the yields of
$(\Lambda+\Sigma^0)$ and $\Xi^-$ at midrapidity from 5\% central
Au+Au collisions as a function of the invariant energy
$\sqrt{s_{NN}}$ in comparison to the available data from Refs.
\cite{exp1b,exp2c}.
We recover a "horn" structure, similar to that shown in Fig.
\ref{horn} for the energy dependence of the strange to non-strange
particle ratios. A sensitivity on the nuclear model parametrizations
persists at low energy, while in the top SPS energy regime the
results corresponding to the different scenarios merge. The
comparison with the available data at $\sqrt{s_{NN}}<8\,$GeV
supports the validity of the CSR picture, while at larger energies
we under-estimate the experimental observations. We mention that
this discrepancy is not due to the CSR mechanism, since it does not
play an essential role in the high-energy regime as pointed out
above.

\section{System size and centrality dependence of strangeness production}
\begin{figure}[t!]
\centering
\includegraphics[width=0.48\textwidth]{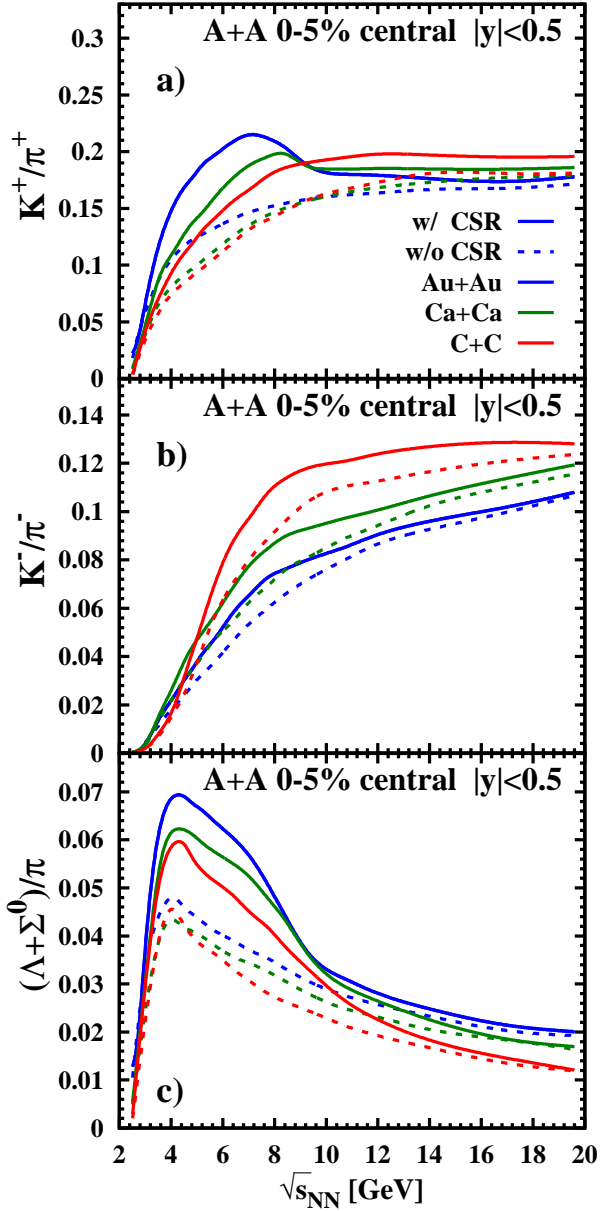}
\vspace{-0.7cm}\caption{(Color online) The ratios $K^+/\pi^+$,
$K^-/\pi^-$ and $(\Lambda+\Sigma^0)/\pi$ at midrapidity from 5\%
central symmetric A+A collisions as a function of the invariant
energy $\sqrt{s_{NN}}$. The solid lines show the results from PHSD
including CSR with NL1 parameters, the dashed lines show the result
from PHSD without CSR. The blue lines refer to Au+Au collisions, the
green lines to Ca+Ca collisions and the red lines to C+C
collisions.}
 \label{horn_systems}
\end{figure}

In this section we explore new aspects of CSR in heavy-ion
collisions. First, we analyze the dependence of the strange to
non-strange particle ratios on the size of the colliding system (cf.
also  Ref. \cite{Cleymans}). Second, we investigate the effects of
CSR on the strange particle yields for different centralities of
Au+Au collision. In Fig. \ref{horn_systems} we present the particle
ratios $K^+/\pi^+$, $K^-/\pi^-$ and $(\Lambda+\Sigma^0)/\pi$ from
PHSD for three types of collision systems. The aim here is to
explore how the variation of the system size modifies the excitation
functions shown in the previous section. In Fig. \ref{horn_systems}
we display the calculations for $^{197}$Au + $^{197}$Au in blue, for
$^{40}$Ca + $^{40}$Ca in green and for $^{12}$C + $^{12}$C in red.
The  scenarios considered are the default PHSD without CSR (dashed
lines) and PHSD including CSR with NL1 as parameter set for the
nuclear equation of state from the non-linear $\sigma-\omega$ model
(solid lines). The inclusion of  CSR gives a strangeness enhancement
also in case of smaller system size with respect to Au+Au collisions
and this holds for all three particle ratios. In fact, when
considering central collisions, a sizeable volume of the system is
affected by the partial restoration of chiral symmetry even in case
of light ions. We notice that, for the $K^+/\pi^+$, $K^-/\pi^-$
ratios, the discrepancy between the calculations with and without
CSR remains sizable even at high SPS energies for Ca+Ca and C+C
collisions. In particular the spread between the scenarios with and
without CSR is larger when the size of the system is smaller. This
can be explained by the fact that in Ca+Ca and C+C collisions the
fraction of the system, which performs the phase transition to the
QGP, is smaller with respect to  Au+Au collisions, and the string
excitations and decays still have a large strangeness production
rate even at larger energies.

These characteristics are evident also in the observation that at
large energies the ratio $K^+/\pi^+$ is smaller for the Au+Au
collisions and larger in C+C collisions. In fact, we recall that the
drop of the $K^+/\pi^+$ ratio in Fig. \ref{horn} is due to the
appearance of the QGP, since the strangeness production in the QGP
phase is suppressed with respect to the hadronic production at fixed
energy density. Concerning the "horn" structure in the $K^+/\pi^+$
ratio, we notice that the peak of the excitation function becomes
less pronounced in case of Ca+Ca and it disappears completely in
case of C+C collisions. With decreasing system size  the low energy
rise of the excitation functions becomes less pronounced. We can see
also that the peak for Ca+Ca is shifted to larger energies with
respect to the Au+Au case. Differently from the $K^+/\pi^+$, the
$(\Lambda+\Sigma^0)/\pi$ ratio preserves the same structure for all
three colliding systems. In order to produce $\Lambda$`s the
threshold energy of $\sqrt{s_{th}}=2.55\,$GeV (for $\Sigma^0$
$\sqrt{s_{th}}=2.62\,$GeV) must be reached, so the
$(\Lambda+\Sigma^0)/\pi$ ratio increases when the system easily
exceeds this value. The peak of the $\Lambda$ production is not
exactly in correspondence of the threshold energy, since we are
considering A+A collisions where the available collision energy is
distributed among participants and where secondary and even higher
order interactions take place. However, it is interesting to notice
that the peak position in this excitation function does not move for
different systems, different from the $K^+/\pi^+$ ratios. At large
energies the $(\Lambda+\Sigma^0)/\pi$ ratio decreases as a function
of the energy, since the pion production is enhanced in the hadronic
re-scattering. Finally, we observe no peak structure in the energy
dependence of the $K^-/\pi^-$ ratio in any of the scenarios studied.
We notice that the results for the different sizes of the system
present an opposite hierarchy with respect to the $K^+/\pi^+$ and
the $(\Lambda+\Sigma^0)/\pi$ ratios. In fact, for C+C and Ca+Ca
collisions the pion production is suppressed, since in the small
systems the hadronic re-scattering cannot develop as in Au+Au
collisions.

\begin{figure}[t!]
\centering
\includegraphics[width=0.45\textwidth]{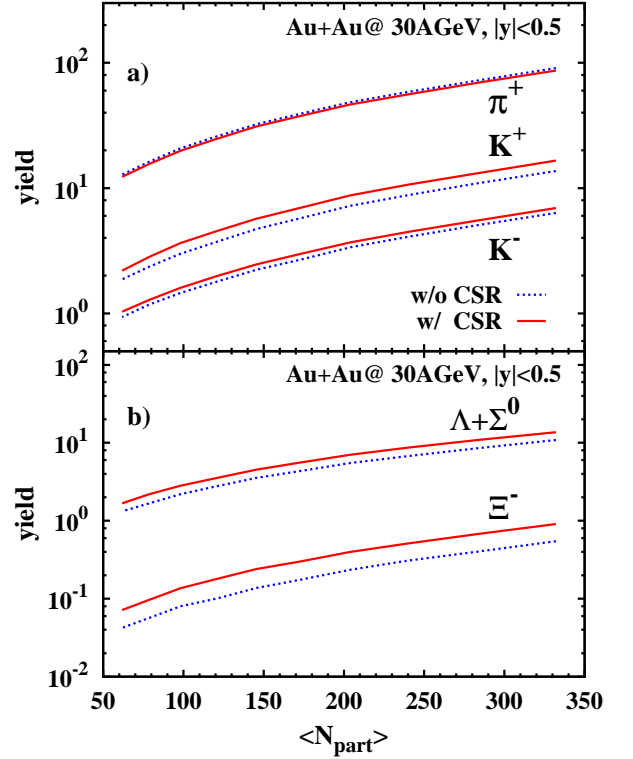}
\vspace{-0.7cm}\caption{(Color online) The particle yields of
$\pi^+$, $K^+$, $K^-$, $\Lambda+\Sigma^0$ and $\Xi^-$ at midrapidity
from Au+Au collisions at 30\,AGeV as a function of the number of
participants. The solid (red) lines show the results from PHSD
including CSR with NL3 parameters, the dotted (blue) lines result
from PHSD without CSR.}
 \label{E30yields}
\end{figure}

\begin{figure}[h!]
\centering
\includegraphics[width=0.45\textwidth]{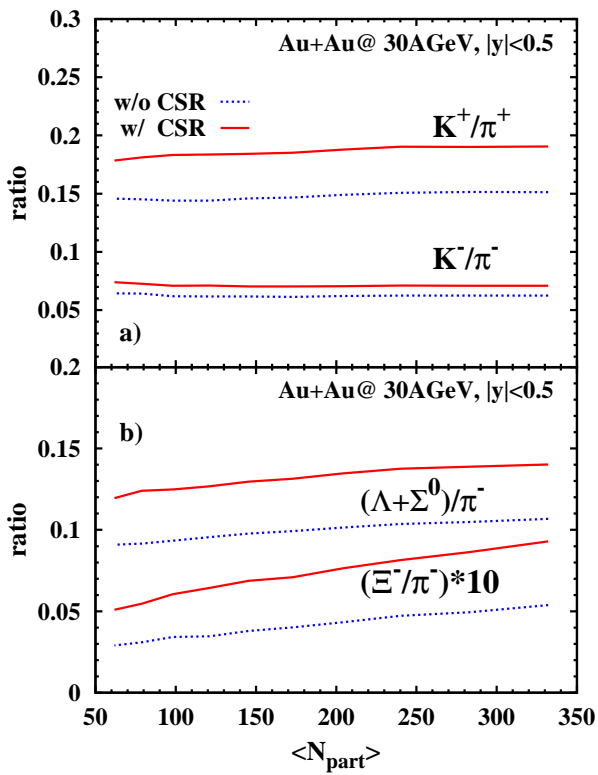}
\vspace{-0.7cm}\caption{(Color online) The particle ratios of
$K^+/\pi^+$, $K^-/\pi^-$, $(\Lambda+\Sigma^0)/\pi^-$ and
$\Xi^-/\pi^-$ (the last one increased by a factor of 10) at
midrapidity from 5\% central Au+Au collisions at 30\,AGeV as a
function of the number of participants. The coding of the lines is
the same as in Fig. \ref{E30yields}.}
 \label{E30ratios}
\end{figure}

Furthermore, in Fig. \ref{E30yields} the abundances of pions, kaons
and the most abundant hyperons are plotted as a function of the
number of participants $\langle N_{part} \rangle$ at midrapidity
from  Au+Au collisions at 30\,AGeV and in Fig. \ref{E30ratios} the
ratios $K^+/\pi^+$, $K^-/\pi^-$, $(\Lambda+\Sigma^0)/\pi^-$ and
$\Xi^-/\pi^-$ are shown for the same collision configuration. In
both cases we show the calculations from PHSD including CSR with NL3
as parameter set by solid red lines and the calculations from PHSD
without CSR by dotted blue lines.

All particle yields decrease with decreasing number of participants.
On the other hand, the ratios appear to be almost constant as a
function of the centrality for $\langle N_{part}\rangle >$ 50. Only
the $\Xi^-/\pi^-$ ratio smoothly decreases with decreasing $\langle
N_{part}\rangle$. The inclusion of CSR produces a strangeness
enhancement in the whole range of centralities  investigated. Note,
however, that very peripheral reactions are not considered for
$\langle N_{part}\rangle >$ 50. At $E_{Lab}=30\,$AGeV the
interaction volume of the two colliding nuclei reaches high energy
densities such that practically all central cells are influenced by
the CSR mechanism independently of the centrality of the collision.
Future heavy-ion collision experiments are expected to shed further
light on the dynamics of the chiral symmetry restoration by
exploring these kind of additional observables.

\section{Thermodynamical aspects of strangeness production in central HIC}
The aim of this section is to study which parts of the phase diagram
in the ($T, \mu_B$)-plane are probed by heavy-ion collisions with
special focus on the strangeness production. In general it is not
straight forward (or even impossible)  to connect non-equilibrium
dynamics from microscopic transport studies to macroscopic
equilibrium properties like temperature and chemical potentials. For
this purpose one needs the exact QCD-equation of state that relates
the energy and the conserved charges to temperature and the chemical
potentials of an equilibrated system. As long as lattice
calculations at finite chemical potential are prevented by the
sign-problem one has to rely on effective models, making a study of
the phase diagram model dependent. Another issue is the question if
the system reaches a local equilibrium during the heavy-ion
collision. A common method to decide on kinetic equilibration is the
pressure equilibration of the energy momentum tensor $T^{\mu \nu}$.
In the local rest-frame it takes the form
\begin{equation} \label{eq:tmunu}
 T^{ \mu \nu}=\begin{pmatrix}
             \epsilon & 0 & 0 & 0 \\
             0 & P_{x} & 0 & 0 \\
             0 & 0 & P_{y} & 0 \\
             0 & 0 & 0 & P_{z}
            \end{pmatrix}\,,
\end{equation}
where $\epsilon$ is the energy density and $P_x$, $P_y$ and $P_z$
are the pressure components in $x$-, $y$- and $z$-direction. In the
center of the collision they are often labeled as
$P_x=P_y=P_{\perp}$ and $P_z=P_{\|}$, when the beam is in the
$z$-direction. Due to the initial asymmetry of the collision the
longitudinal and the transverse pressure differ significantly. A
necessary requirement for kinetic equilibrium is the coincidence of
the pressure components $P_{\perp} \approx P_{\|}$. The behavior of
the pressure components in the central region of the collision zone
has been studied in Ref. \cite{Bravina:1999kd} with the UrQMD
transport model \cite{Bass:1998ca,Bleicher:1999xi}. It was found
that the pressure equilibrates at $t \cong 10 \,$fm/c after the
initial impact of central Au+Au collisions at AGS energies.
Additionally, a good agreement between the energy spectra of
different hadron species with the predictions of statistical models
was found at this time. This indicates that one can indeed find an
equilibrated system at AGS energies for times larger than $\sim$ 10
fm/c.

To fix points in the ($T, \mu_B$)-plane we have to determine the
temperature $T$ and the baryon chemical potential $\mu_B$ of the
medium in the expanding fireball. This is usually done by comparing
the energy density and the conserved charges in the local cell to
the corresponding equation of state \cite{Endres:2015egk}. For
hadronic matter it is common to use a hadron-resonance gas  equation
of state. However, it is unclear which hadronic resonances should be
included in such a model. We will therefore determine a temperature
$T$ and a baryon chemical potential $\mu_B$ from the energy density
and particle density of nucleons and pions which are directly
accessible within our transport simulations. Instead of examining
the whole fireball we will focus on local cells where strangeness is
produced; these cells may be close to thermal equilibrium (at late
times) or out-of equilibrium (at early times). The focus on local
cells with strangeness production excludes free streaming cells as
well as everything that happens after chemical freeze-out. Whenever
a new $s\bar{s}$-pair is produced in the hadronic medium we take the
nucleon and pion energy densities $\epsilon_N$ and $\epsilon_{\pi}$ and determine
the temperature $T$ and the baryon chemical potential $\mu_B$ using
the expressions for a noninteracting hadron gas (in equilibrium):
\begin{align}
 \rho_{\pi} &=g_{\pi} \int \frac{\mathrm{d}^3 p}{(2 \pi)^3} \frac{1}{\mathrm{e}^{\omega_{\pi}/T}-1}\,,\\
 \epsilon_{\pi} &=g_{\pi} \int \frac{\mathrm{d}^3 p}{(2 \pi)^3} \frac{\omega_{\pi}}{\mathrm{e}^{\omega_{\pi}/T}-1}\,,\\
 \rho_{N} &=g_N \int \frac{\mathrm{d}^3 p}{(2 \pi)^3} \frac{1}{\mathrm{e}^{(\omega_{N}-\mu_B)/T}+1}\,,\\
 \epsilon_{N} &=g_N \int \frac{\mathrm{d}^3 p}{(2 \pi)^3} \frac{\omega_N}{\mathrm{e}^{(\omega_{N}-\mu_B)/T}+1}\,,
\end{align}
where $\omega_i=\sqrt{{\bf p}^2+m^2_i}$ is the energy of the
respective particle and $g_{\pi}=3$ and $g_N=4$ are the degeneracy
factors of pions and nucleons. By additionally evaluating the
densities $\rho_{\pi}$ and $\rho_{N}$ we can check if the
local cell is in approximate thermal equilibrium or not. In
practical terms: If the temperatures obtained from both methods
differ by more than $5\,$MeV or if the chemical potentials differ by
more than $15\,$MeV we consider the cell to be out-of equilibrium.
Within this procedure we can eliminate further cells from the phase
diagram, which are out-of equilibrium, however, it does not change
the probed area if one considers only events that happen $10\,$fm/c
after the collision.
\begin{figure}[t!]
   \centering
   \includegraphics[width=1.0\linewidth]{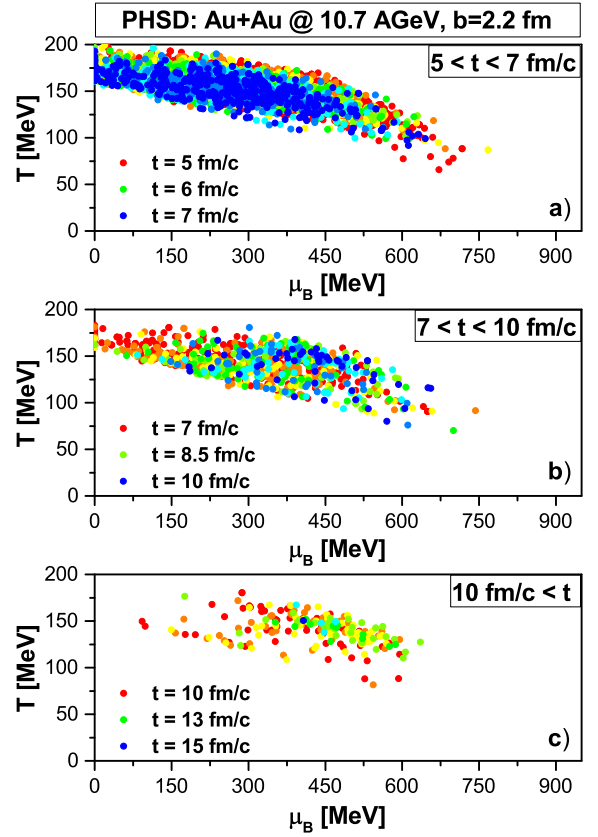}
   \caption{(Color online) Occupancy of the phase diagram for hadronic matter in a central Au+Au collision
   at $10.7\,$AGeV for different time intervals. Each point belongs to a cell where strange quarks were produced.
   The color of the points indicates the time of the events within some varying interval.
   For times $>$ 10 fm/c the strangeness production occurs in cells that are in approximate thermodynamic equilibrium
   while the cells in panels (a) and (b) are dominantly out-of equilibrium.}
   \label{pic:pd_vs_time}
\end{figure}

\begin{figure}[t!]
   \centering
   \includegraphics[width=1.0\linewidth]{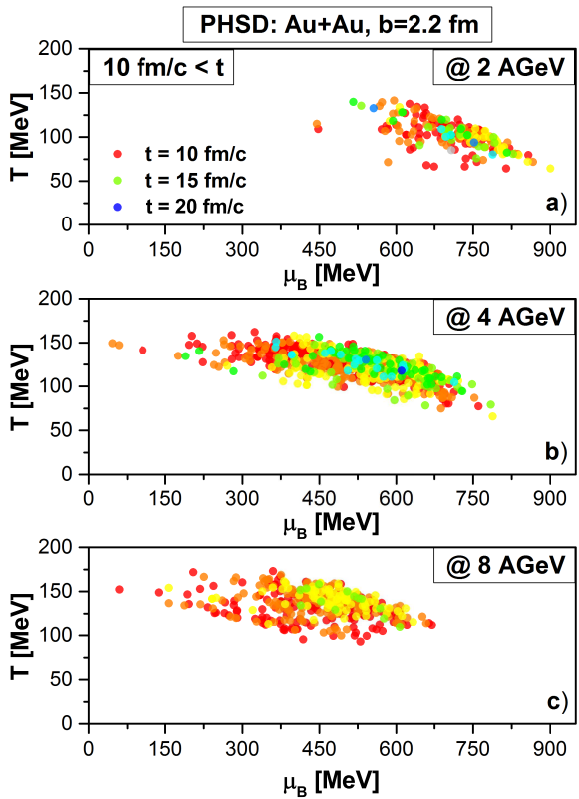}
   \caption{(Color online) Occupancy of the phase diagram for hadronic matter in Au+Au collisions at different beam energies from 2 to 8 AGeV for
   times $t > $ 10 fm/c. Each point belongs
to a cell where strange quarks were produced. The color of the points indicates the time of the events within some varying interval.}
   \label{pic:pd_vs_energy}
\end{figure}

Fig. \ref{pic:pd_vs_time} shows the reconstructed temperatures and
chemical potentials for a Au+Au collision at $10.7\,$AGeV with an
impact parameter $b=2.2\,$fm. This reaction is clearly dominated by
local cells with partonic content since hadrons in QGP cells
dissolve since the energy density is above critical. Each point in
Fig. \ref{pic:pd_vs_time} stands for a local cell in which new
strange quarks where produced. The color of the points indicates the
time of the production (red points early, yellow and green points at
intermediate times, blue points at the end of the time interval).
The panel (a) shows the events from $5 - 7\,$fm/c after the initial
collision. These points cover all chemical potentials up to
$\mu_B=750\,$MeV. The maximum temperature is $T=200\,$MeV when
adopting only baryons and pions as degrees-of-freedom. These cells
in panel (a) are dominantly out-of thermodynamic equilibrium  and
for low $\mu_B$ and high $T$ correspond to partonic cells with some
pion content.  The panel (b) shows the events from $7 - 10\,$fm/c
where the majority of the points are located at baryon chemical
potentials larger than $\mu_B=150\,$MeV. The blue points in panel
(b) belong to cells that are already in equilibrium. The panel (c)
shows the events that happen at times $t >10\,$fm/c after the
initial collision and are approximately  in equilibrium. They cover
an area between $150\,\text{MeV}<\mu_B<650\,\text{MeV}$ and
$100\,\text{MeV} < T < 175\,\text{MeV}$. The panel (c) describes
indeed a proper phase diagram while the upper two (a) and (b) suffer
from non-equilibrium effects. We stress again that the temperatures
and baryon chemical potentials that are shown in the panels (a) and
(b) can not be related to any equilibrium properties. In spite of
the restrictions imposed by thermodynamic equilibrium we find that
even central Au+Au collisions at 10.7 AGeV explore a wide range of
points in the $T, \mu_B$-plane for $\mu_B$ essentially below 650
MeV.

Fig. \ref{pic:pd_vs_energy}, furthermore, shows the occupation of
the phase diagram extracted from Au+Au collisions at $2$ (a), $4$
(b) and $8\,$AGeV (c) with an impact parameter $b=2.2\,$fm. All
points belong to events that happened at times $t > 10\,$fm/c after
the initial collision. One sees that the probed region shifts to
larger baryon chemical potentials and smaller temperatures when
lowering the beam energy. The maximum baryon chemical potential in
these plots is $\mu_B=900\,$MeV, the lowest temperature $T=65\,$MeV,
however, with a very large spread in $T$ and $\mu_B$.

It is important to discuss how these occupancies in the phase
diagram relate to the real QCD-phase diagram. Due to the model
dependent EoS the extracted temperatures and chemical potentials do
not represent the real ones for QCD. We recall that the PHSD
transport approach uses a critical energy density of
$\epsilon_c=0.5\,$GeV/fm$^3$ to distinguish between a hadronic and a
partonic medium. If the local energy density is above this threshold
the hadrons dissolve into quarks. The critical energy density
$\epsilon_c$ marks the largest energy density a hadronic system can
reach in our simulations. When compared to recent lattice results
from the Wuppertal-Budapest collaboration \cite{Borsanyi:2013bia} at
$\mu_B=0$ this translates to a temperature of around $T \approx
160\,$MeV. On the other side the model used to extract the
temperature and chemical potential is based on the hadron resonance
gas with a reduced number of degrees-of-freedom. We mention that a
Hadron Resonance Gas (HRG), that contains all the hadronic particles
included in PHSD \footnote[3]{The hadronic particles included in
PHSD are the 0$^-$ and 1$^-$ meson octets, the spin 1/2 and 3/2
baryon octets, the N(1440)and N(1535) resonances and the a$_1$
meson.}, reaches the critical energy density at a temperature $T
\approx 175\,$MeV. This indicates that the temperatures shown in
Figs. \ref{pic:pd_vs_time} and \ref{pic:pd_vs_energy} are too large
compared to full QCD. As a rough guide one should divide the
temperatures in these figures by a factor $\sim$ 1.1 in order to
obtain an estimate closer to full QCD. For the baryon chemical
potential we note that the baryon number susceptibilities $\chi_B$
of the HRG are smaller than the lattice results
\cite{Borsanyi:2013bia}. This implies the corresponding baryon
densities, in first order given by $n_B \approx \chi_B \ \mu_B$,
exceed the HRG densities, thus overestimating the extracted baryon
chemical potentials $\mu_B$ in comparison to full QCD. Admittedly we
can not give a definite rescaling for finite chemical potentials,
nevertheless, the general trend should be the same in the whole
$T-\mu_B$ -plane shifting the probed area to smaller temperatures
and chemical potentials. Nevertheless, it becomes apparent from Fig.
\ref{pic:pd_vs_energy} that it will be very hard to identify a
critical point in the ($T, \mu_B$)-plane experimentally since the
spread in $T$ and $\mu_B$ is very large at all bombarding energies
of interest.

\section{Summary}
In this work we have analyzed the effects of chiral symmetry
restoration (CSR) on  observables from heavy-ion collisions in the
energy range $\sqrt{s_{NN}}=3-20\,$GeV in extension of the
earlier study in Ref. \cite{PHSD_CSR}. Our results have been
obtained within the Parton-Hadron-String Dynamics transport approach
\cite{PHSD}, where essential aspects of CSR have been incorporated
in the Schwinger mechanism for the string decay \cite{PHSD_CSR}.
Since the PHSD approach includes both hadronic and partonic
degrees-of-freedom and has been tested in a wide energy regime, it
represents a powerful tool to study nucleus-nucleus collisions on a
microscopic basis. The CSR, as implemented in PHSD, affects only the
hadronic particle production and it does not imply modifications  in
the Quark-Gluon Plasma (QGP) phase. As already found in Ref.
\cite{PHSD_CSR} the CSR induces an enhancement of the strange quark
fraction $\gamma_s$ produced via the string decay, while there are
no sensible changes in the diquark production and accordingly in
baryon-antibaryon production. The $s/u$ ratio, as defined by the
Schwinger formula (\ref{Schwinger-formula}), increases as a function
of the energy density due to CSR and this is reflected in an
enhancement of the strange particle abundances with respect to the
non-strange ones. This has been observed explicitly in the particle
spectra at AGS and lower SPS energies ($E_{Lab}=10.7$ to 30\,AGeV).
On the other hand, at top SPS energies (e.g. $E_{Lab}=158\,$AGeV)
the results from PHSD with and without CSR merge, since the dynamics
of the system is dominated by the QGP phase where CSR does not play
a significant role. 

In extension to Ref. \cite{PHSD_CSR} we
have performed calculations for different nuclear equations of state
(NL1, NL2 and NL3) and thus could quantify the uncertainties in the
particle yields and ratios (cf. Figs. 6-8 and 13, 14 as well as table II). Since the sets
NL1 and NL2 give a larger scalar nucleon density $\rho_s$ at the same
energy density as the set NL3 the "horn" in the $K^+/\pi^+$ is more
pronounced and closer to the experimental data. Furthermore, we
found that the transverse mass spectra are only slightly modified by
the inclusion of the CSR and practically insensitive to the nuclear
EoS. In fact, the CSR mechanism acts predominantly on the chemistry
and not on the kinematics of the string decays.

We stress that our PHSD calculations provide a microscopic
interpretation of the "horn" structure in the excitation function of
the $K^+/\pi^+$ ratio in central Au+Au (or Pb+Pb) collisions. The
steep rise of this ratio at AGS energies is associated to CSR, while
the drop at higher SPS energies is due to the appearance of the QGP
phase in an increasing volume of the interaction region. We have
found an analogous energy dependence for the
$(\Lambda+\Sigma^0)/\pi$ ratio, while the excitation function of the
$K^-/\pi^-$ ratio does not show any explicit peak.
In general, the PHSD results
obtained with the inclusion of CSR are in a good agreement with the
available data for all observables analyzed, while calculations
without CSR fail substantially.

In extension to Ref. \cite{PHSD_CSR} we have investigated also
different sizes of the colliding ions ($^{197}$Au, $^{40}$Ca and
$^{12}$C) and computed the strange to non-strange particle ratios
for these configurations with and without CSR. It is found that  the
"horn" feature in the $K^+/\pi^+$ ratio appears only for larger
system sizes, i.e. Au+Au and Ca+Ca, while the "horn" disappears in
case of C+C collisions. The $(\Lambda+\Sigma^0)/\pi$ excitation
function maintains the peak-structure as we have observed in case of
Au+Au collisions also for smaller sizes of the system. Furthermore,
we have analyzed the strange particle abundances in Au+Au collisions
at $30\,$AGeV as a function of the number of participants in the
collision. As mentioned above, when including CSR in the PHSD
calculations we obtain an increase of the strange particle yields
with respect to the results from PHSD without CSR and this feature
is valid in case of central collisions as well as moderate
peripheral collisions. More experimental observations are needed to
extract information about the centrality and system size dependence
of the CSR.

We have, furthermore, addressed the question whether the strangeness
production in HIC occurs in thermodynamical equilibrium or not and
have found that strange particles are produced dominantly at the
early stages of the collisions, when the system is not in thermal
and chemical equilibrium. At AGS energies only a few percent of the
total strangeness production happens in approximate local
thermodynamical equilibrium. With decreasing bombarding energy lower
temperatures and higher baryon chemical potentials are reached,
however, the spread in $T$ and $\mu_B$ is very large such that a
search for a critical point in the phase diagram becomes very
difficult experimentally.

In conclusion, our microscopic studies support the idea that CSR
occurs in hadronic systems with high temperatures and densities
before the deconfinement phase transition takes over. We suggest
that the strange particle spectra and yields are suitable signatures
to study the properties of CSR in HICs in future also as a function
of system size and centrality.

\section*{Acknowledgments}
The authors acknowledge inspiring discussions with J. Cleymans, M.
Gazdzicki, M. Gorenstein and O. Linnyk and thank the Helmholtz
International Center for FAIR (HIC for FAIR), the Helmholtz Graduate
School for Hadron and Ion Research (HGS-HIRe), the Helmholtz
Research School for Quark Matter Studies in Heavy-Ion Collisions
(H-QM), and the Bundesministerium f\"{u}r Bildung und Forschung
(BMBF) for support. The computational resources have been provided by the Center for Scientific Computing (CSC) in the framework of the LandesOffensive zur Entwicklung Wissenschaftlich-\"{o}konomischer Exzellenz  (LOEWE).

\end{document}